\documentclass{scrbook}

\usepackage{amsmath}
\usepackage{cite}
\usepackage{graphicx}

\DeclareMathOperator{\tr}{tr} 

\title{Excited-state calculations with quantum Monte Carlo}

\author{Jonas Feldt and Claudia Filippi \\\\
\textit{MESA+ Institute for Nanotechnology, University of Twente,}\\
\textit{P.O. Box 217, 7500 AE Enschede, The Netherlands}
}

\date{August 2019}
\publishers{to be published in ``Quantum Chemistry and Dynamics of Excited States: Methods and Applications'',\\
Eds. L.\ Gonz\'alez and R.\ Lindh, 2020 Wiley-VCH}

\begin{document}

\maketitle

\noindent
\textbf{\Large Abstract}\\[3ex]

\noindent
Quantum Monte Carlo methods are first-principle approaches that approximately solve
the Schr\"odinger equation stochastically. As compared to traditional quantum
chemistry methods, they offer important advantages such as the ability to handle a
large variety of many-body wave functions, the favorable scaling with the number
of particles, and the intrinsic parallelism of the algorithms which are particularly
suitable to modern massively parallel computers.  In this chapter, we focus on
the two quantum Monte Carlo approaches most widely used for electronic structure
problems, namely, the variational and diffusion Monte Carlo methods.  We give
particular attention to the recent progress in the techniques for the optimization 
of the wave function, a challenging and important step 
to achieve accurate results in both the ground and the excited state.
We conclude with an overview of the current status of excited-state calculations 
for molecular systems, demonstrating the potential of quantum Monte Carlo methods
in this field of applications.

\chapter{Excited-state calculations with quantum Monte Carlo}

\section{Introduction}
\label{sec:intro}

Quantum Monte Carlo (QMC) methods are a broad range of approaches which employ
stochastic algorithms to simulate quantum systems, and have been used to study
fermions and bosons at zero and finite temperature with very different many-body
Hamiltonians and wave functions in the fields of molecular chemistry, condensed
matter, and nuclear physics.  While all QMC methods, despite the diversity of
applications, share some core algorithms, we restrict ourselves here to the two
zero-temperature continuum QMC methods\footnote{A quantum Monte Carlo approach not 
in real space but in Slater determinant space (i.e.\ the full configuration interaction QMC method)
is briefly introduced in Chapter 1.} that are most commonly used in electronic
structure theory, namely, variational (VMC) and diffusion (DMC) Monte 
Carlo~\cite{Foulkes2001,luchow_quantum_2011,austin_quantum_2012}.

\begin{figure}
	\includegraphics[width=\linewidth]{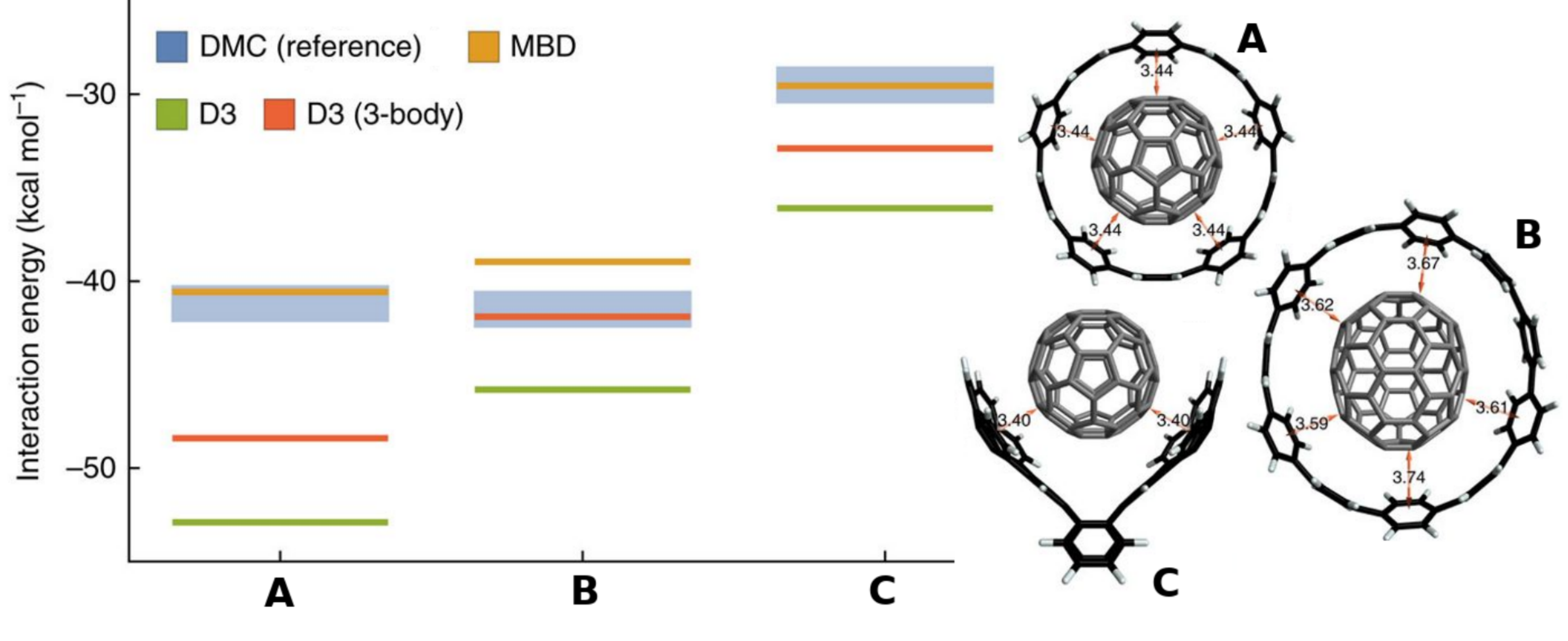}
	\caption{Interaction energies of the complexes \textbf{A}--\textbf{C} computed with
        DMC and density functional theory with two different dispersion corrections (D3) and 
        the many-body dispersion method (MBD). Adapted from Ref.~\cite{hermann_nanoscale_2017}.}
	\label{fig:intro}
\end{figure}

As compared to deterministic quantum chemistry approaches, solving the
Schr\"odinger equation by stochastic means in VMC or DMC offers several
advantages.  The stochastic nature of the integration allows for a greater
flexibility in the functional form of the many-body wave function employed,
which can for instance include the explicit dependence on the inter-electronic
distances. As a consequence, more compact wave functions can be used (the number
of determinants needed to get the same energy is reduced by a few orders of
magnitude) and, further, the dependence on the basis set is much weaker.  Even
though VMC and DMC are expensive, they have a favorable scaling with the system
size (a mere polynomial $N^4$ in the number of electrons $N$), which has enabled
simulations with hundreds and even thousands of electrons also in condensed
matter, where traditional highly-correlated approaches are very difficult to
apply. Finally, the intrinsically parallel nature of QMC algorithms renders
these methods ideal candidates to take advantage of the massively parallel
computers which are now available. An impressive example of such calculations 
is shown in Fig.~\ref{fig:intro} where the interaction energy dominated by 
dispersion is benchmarked with DMC for remarkably large complexes~\cite{hermann_nanoscale_2017}.

That said, when inspecting the literature, it is evident that VMC and DMC
methods have traditionally been employed to calculate mainly total energies and
total energy differences as the computation of quantities other than the energy
is more complicated. QMC calculations are for instance generally carried out on
geometries obtained at a different level of theory and also the construction of
the many-body wave function and its optimization are not straightforward. 
Particular care must in fact be paid to this step since the residual DMC error
can be larger than sometimes assumed in the past, when calculations were anyhow
limited to relatively simple wave functions and it was not
feasible to extensively explore the dependence of the results on the choice of
wave function. We will come back below to this point, which is
especially relevant for excited states.

The last few years have however seen remarkable progress in methodological
developments to overcome these and other limitations, as well as extend the
applicability of QMC to larger systems both in the ground and the excited state.
In particular, robust optimization algorithms for the parameters in the wave
function have been developed for ground states~\cite{Umrigar2005, Umrigar2007,
    Toulouse2007a,Neuscamman2017} and extended to the
state-average~\cite{filippi_absorption_2009} as well as
state-specific~\cite{zhao_efficient_2016, shea_size_2017} optimization of
excited states. Importantly, it has recently become possible to efficiently
compute the quantities needed in these optimization schemes (i.e.\ the
derivatives of the wave function and the action of the Hamiltonian on these
derivatives) at a cost per Monte Carlo step which scales
like the computation of the energy alone~\cite{Sorella2010, Filippi2016,
   assaraf_optimizing_2017}.  Consequently, the determinantal component of a QMC
wave function does not have to be borrowed from other quantum chemical
calculations but can be consistently optimized within VMC after the addition of
the correlation terms depending on the inter-electronic distances. These
developments also enable the concomitant optimization of the structural
parameters in VMC even when large determinantal expansions are employed in the
wave function~\cite{assaraf_optimizing_2017,dash_perturbatively_2018}.  
The possibility of performing molecular dynamics simulations with VMC forces has also been
recently demonstrated~\cite{Sorella2014, Sorella2015}.  

Researchers have also been actively investigating more complex functional forms
~\cite{Lopez2006,holzmann_orbital-dependent_2019,Casula2003,Casula2004,Bajdich2008} 
to recover missing correlation and allow a more
compact wave function than the one obtained with a multi-determinant
component.  A local correlation description has also been shown to be a promising route to achieve 
smaller expansions and reduced computational costs for ground and excited
states~\cite{fracchia_size-extensive_2012, Zulfikri2016}.
In parallel, algorithms have been
explored for a more automatic selection of the determinantal component,
avoiding the possible pitfalls of a manual choice based on chemical
intuition~\cite{clay_influence_2015,
per_energy-based_2017, robinson_excitation_2017,kim_qmcpack_2018, 
dash_perturbatively_2018,scemama_excitation_2018, flores_excited_2018}. 
Importantly, effort has been devoted to develop algorithms for the computation of
quantities other than the energy via estimators characterized by reduced
fluctuations as well as wave function bias~\cite{Assaraf1999, Nightingale2001, Assaraf2003, Umrigar2005,
    Toulouse2007b, Attaccalite2008, per_efficient_2012}.
In addition to these methodological advances, various tools have become available to
facilitate the calculations, such as tables of pseudopotentials and corresponding basis sets
especially constructed for QMC~\cite{Burkatzki2007, Burkatzki2008,Trail2017,
    Mitas2017, Mitas2018a, Mitas2018b}.
Finally,
multi-scale methods have been proposed to include the effects of a (responsive)
environment on an embedded system treated with QMC~\cite{floris_density_2012, floris_electronic_2014,
    daday_wavefunction_2014, guareschi_solvent_2014, guareschi_introducing_2016,
    doblhoff2018}.

After a brief description of the VMC and DMC methods, we will focus here 
on some of these recent developments, giving special attention to the 
algorithms employed to optimize the variational parameters in the wave
function. We will then review relevant work and recent advances in the
calculation of excited states and their properties, mainly for molecular
systems. 
We note that useful sources for QMC are the introductory book to Monte
Carlo methods and their use in quantum chemistry~\cite{Hammond1994}, and the
reviews on QMC methods and their applications to solids~\cite{Foulkes2001,
    kolorenc_applications_2011, Wagner2016, Wagner2018} and to noncovalent
interactions~\cite{dubecky_noncovalent_2016}. A detailed introduction to VMC and
DMC can be found in Ref.~\cite{toulouse_chapter_2016}. Finite-temperature path
integral Monte Carlo methods are covered in Ref.~\cite{Ceperley1995}.

\section{Variational Monte Carlo}
\label{sec:vmc}

Variational Monte Carlo is the simplest flavor of QMC methods and represents a
generalization of classical Monte Carlo to compute the multidimensional
integrals in the expectation values of quantum mechanical operators.  The
approach enables the use of any ``computable'' wave function without severe
restrictions on its functional form. This must be contrasted to other
traditional quantum chemical methods which express the wave function as products
of single particle orbitals in order to perform the relevant integrals
analytically.

To illustrate how to compute an expectation value stochastically,  let us consider the variational 
energy $E_\mathrm v$, namely, the expectation value of the Hamiltonian $\hat H$ on a given wave function 
$\Psi$, which we rewrite as
\begin{equation}
   E_\mathrm v = \frac{\int \Psi(\mathbf R)^* \hat H \Psi(\mathbf R) \mathrm d\mathbf R}
              {\int |\Psi(\mathbf R)|^2 \mathrm d\mathbf R}
    = \frac{\int |\Psi(\mathbf R)|^2 \frac{\hat H \Psi(\mathbf R)}{\Psi(\mathbf R)} \mathrm d\mathbf R}
              {\int |\Psi(\mathbf R)|^2 \mathrm d\mathbf R}
    = \int \rho(\mathbf R) E_\mathrm L(\mathbf R) \mathrm d\mathbf R\,,
\label{eq:vmc}
\end{equation}
where we have introduced the probability distribution, $\rho(\mathbf R)$, and the local energy, 
$E_\mathrm L(\mathbf R)$, defined as
\begin{equation}
    \rho(\mathbf R) = \frac{|\Psi(\mathbf R)|^2}{\int |\Psi(\mathbf R)|^2 \mathrm d\mathbf R} 
    \ \ \ \text{and} \ \ \ 
    E_\mathrm L(\mathbf R) = \frac{\hat H \Psi(\mathbf R)}{\Psi (\mathbf R)}\,,
\label{eq:rho}
\end{equation}
with $\mathbf R$ denoting the $3N$ coordinates of the electrons. We note that we can interpret
$\rho(\mathbf R)$ as a probability distribution since it is always non-negative and integrates to one.

\begin{figure}
    \centering
    \includegraphics[width=0.6\linewidth]{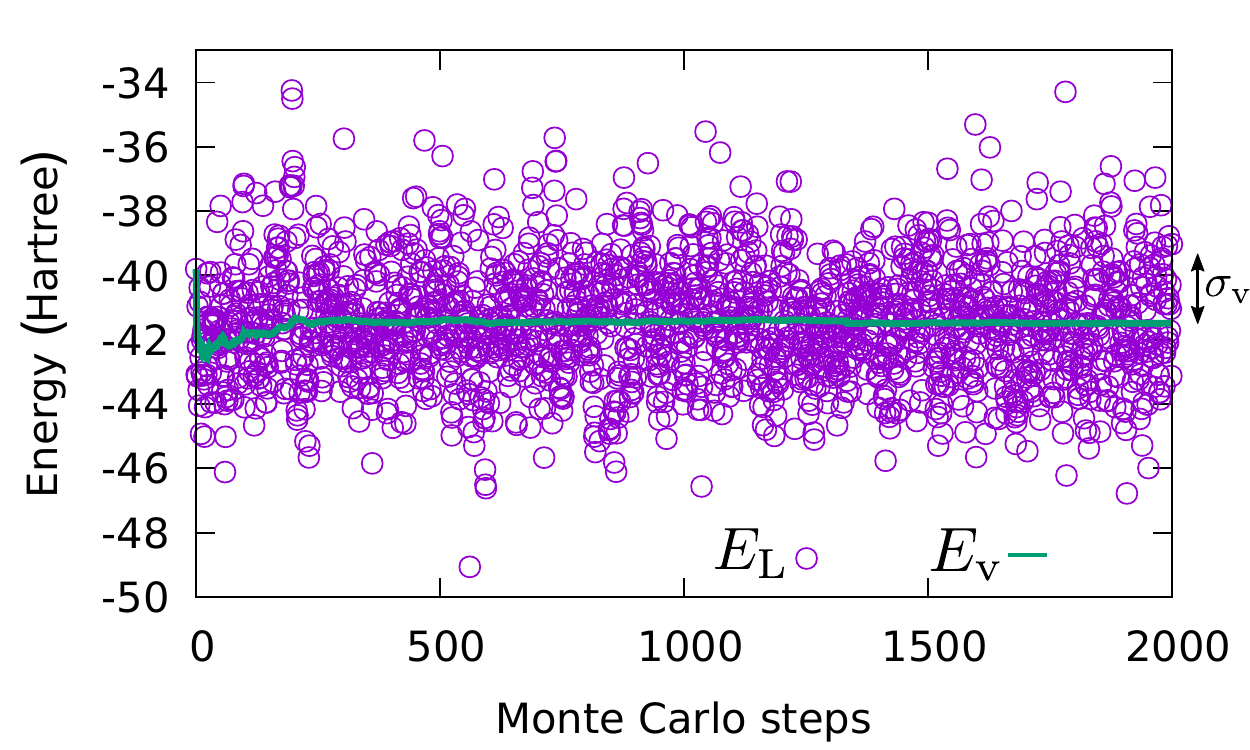}
    \caption{Local energy (circle) and its running average (green line) in a typical VMC run. The size of the root-mean-square fluctuations of the local energy, $\sigma_\mathrm v$, is also indicated.}
\label{fig:vmc}
\end{figure}

The integral can then be estimated by averaging the local energy computed on a set of $M$ configurations
$\{\mathbf R_k\}$ sampled from the probability density $\rho(\mathbf R)$ as
\begin{equation}
    E_\mathrm v \approx \bar{E}_\mathrm L=\frac{1}{M} \sum_k^M E_\mathrm L (\mathbf R_k)\,.
\label{Eave}
\end{equation}
According to the central limit theorem, this estimator converges to the exact value, $E_\mathrm v$, with increasing
number of Monte Carlo configurations with a statistical uncertainty which decreases as
\begin{equation}
    \text{err}(\bar E_\mathrm L) \propto \frac{\sigma_\mathrm v}{\sqrt{M}}\,,
\end{equation}
where $\sigma_\mathrm v^2=\int \rho(\mathbf R) (E_\mathrm L(\mathbf R) -E_\mathrm v)^2 \mathrm d\mathbf R$ 
is the variance of the local energy. For this relation to hold, the chosen wave function must
yield a finite variance of the sampled quantity, in this case, the local energy. A typical VMC run
is illustrated in Fig.~\ref{fig:vmc}, where the local energy is computed at each Monte Carlo step together 
with its running average.

Importantly, the statistical uncertainty decreases as $1/\sqrt{M}$ independently
of the number of dimensions in the problem so that Monte Carlo displays a
faster convergence than deterministic numerical integration already for small
numbers of dimensions\footnote{For instance, the error for the Simpson's integration rule decreases as $1/M^{(4/d)}$ with $d$ the number of dimensions and $M$ the number of integration points, so Monte Carlo integration is more efficient already for $d > 8$. }.
Furthermore, as the trial wave function\footnote{A trial wave function is a
wave function used as an approximation to the state of interest.} improves, the
Monte Carlo estimate of the variational energy requires
fewer Monte Carlo steps to converge.  In the limit of the wave function
being an exact eigenstate of the Hamiltonian, the variance approaches zero and a
single configuration is sufficient to obtain the exact variational energy. This
\textit{zero-variance} principle applies straightforwardly to the Hamiltonian
and operators commuting with the Hamiltonian (therefore, the large number of
total energy calculations found in the QMC literature). This principle can
however be generalized to arbitrary observables by formulating an equivalent,
improved estimator having the same average but a different, reduced
variance~\cite{Assaraf1999}.  Reduced variance estimators have been derived for
the computation of electron density~\cite{Assaraf2007,per_zero-variance_2011}, the
electron pair densities~\cite{Toulouse2007b}, interatomic
forces~\cite{Assaraf2003, Attaccalite2008}, and other derivatives of the total
energy~\cite{Umrigar2005}. 

In practice, the probability distribution $\rho(\mathbf R)$ is sampled with the
Metropolis-Hastings algorithm by simulating a Markov chain. This is a sequence
of successive configurations, ${\mathbf R}_1,\ldots, {\mathbf R}_M$, generated
with a transition probability, $P(\mathbf R'|\mathbf R)$, where the transition
to a new configuration $\mathbf R'$ only depends on the current point $\mathbf
R$. The transition probability is stochastic, which means that it has the
following properties:
\begin{equation}
	P(\mathbf R'|\mathbf R) \geq 0 \ \ \ \text{and} \ \ \ \int P(\mathbf R'|\mathbf R)\mathrm d\mathbf R' = 1\,.
\end{equation}
Repeated application of $P$ generates a Markov chain which converges to the target distribution $\rho$ as
\begin{equation}
\lim\limits_{M\rightarrow \infty} \int P(\mathbf R | \mathbf R_{M})\ldots P(\mathbf R_2 | \mathbf R_1) \rho_\mathrm{init}(\mathbf R_1)\mathrm d \mathbf R_1 \ldots \mathbf R_M= \rho(\mathbf R)\,,
\end{equation}
\textit{if} $P$ is ergodic (it possible to move between two different configurations in a finite number of steps) 
and fulfills the so-called stationarity condition:
\begin{equation}
\int P(\mathbf R' | \mathbf R)\rho(\mathbf R)\mathrm d \mathbf R = \rho(\mathbf R')\,.
\end{equation}
The stationarity condition tells us that, if we start from the desired
distribution $\rho$, we will continue to sample $\rho$. Moreover, if the
stochastic probability $P$ is ergodic, it is possible to show that this
condition ensures that any initial distribution will evolve to $\rho$ under
repeated applications of $P$.

In the Metropolis-Hastings algorithm, the transition to a new state is carried
out in two steps: a new configuration is generated by a (stochastic) proposal
probability and the proposed step is then accepted or rejected with an
acceptance probability. The latter can be constructed so that the combined
proposal and acceptance steps fulfill the stationarity condition. Most
importantly, the acceptance depends only on ratios of $\rho(\mathbf R)$ so that
the generally unknown normalization of the distribution $\rho$ is not required.
We note that it is desirable to reduce sequential correlation among
configurations. Proposing large steps to quickly explore the phase space must
therefore be balanced against the rate of acceptance which decreases with large
steps. For these reasons, electrons are generally moved one at the time to allow
larger steps with a reasonable acceptance rate, a necessary feature as the
system size grows since the size of the move would need to be decreased to have
a reasonable acceptance of a move of all particles.

\begin{figure}
	\centering
	\includegraphics[width=0.8\linewidth]{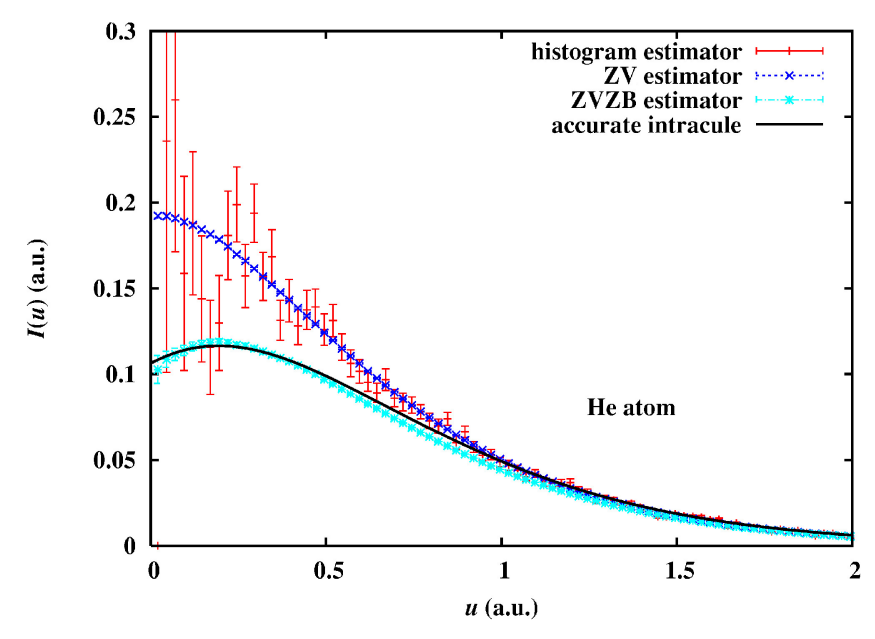}
	\caption{Spherically-averaged intracule density $I(u)$ as a function of the 
        electron-electron distance $u$ for the He atom calculated in VMC with a histogram,
        a zero-variance, and a zero-variance zero-bias estimator and the same
        Hartree-Fock wave function (without a Jastrow factor). Adapted from Ref.~\cite{Toulouse2007b}.}
	\label{fig:zvzb}
\end{figure}

VMC is a very powerful method as the stochastic nature of integration gives a
lot of freedom in the choice of the functional form of the wave function. It also
allows us to learn a great deal about a system by exploring which ingredients in
the wave function are necessary for its accurate description. Finally, in VMC,
there is no sign problem associated with Fermi statistics, which generally
plagues other quantum Monte Carlo approaches as we will see below. The obvious
drawback is that, for each particular problem, a parametrization of the wave function has to be
constructed. This process can be non-trivial and tends to be biased
towards simpler electronic states: for example, it is easier to construct a good
wave function for a closed-shell than an open-shell system so that the energy of
the former will be closer to the exact result than for the latter. Furthermore,
properties other than the energy (or expectation values of operators commuting
with the Hamiltonian) can be significantly less accurate since they are first
order in the error of the wave function instead of second order as for the
energy. It has however been shown that it is possible to extend this favorable 
property of the energy to arbitrary observables by using modified estimators which 
lead not only to reduced fluctuations but also to a reduced bias due to the wave 
function~\cite{Assaraf2003} as convincingly demonstrated in some promising 
applications~\cite{Toulouse2007b,per_efficient_2012, per_zero-variance_2011, Assaraf2003}. 
An example of this so-called zero-variance (ZV) zero-bias (ZB) approach applied to the computation 
of the intracule density is shown in Fig.~\ref{fig:zvzb}: the use of a ZV estimator
significantly reduces the statistical fluctuations of the density and the further 
ZB formulation yields the correct result even when a simple Hartree-Fock wave function is
employed.
In general, the VMC approach is an extremely valuable tool and, in recent years,
its use and impact has in fact become greater thanks to the availability of
robust methods to optimize the many parameters in the wave function and,
consequently, to increase the accuracy of the observables of interest already at
the VMC level. Finally, characterizing and optimizing the trial wave function in
VMC represents a necessary ingredient for more advanced projector Monte Carlo
methods like the diffusion Monte Carlo approach described in next Section.

\section{Diffusion Monte Carlo}
\label{sec:dmc}

Projector Monte Carlo methods are QMC approaches which remove (at least in part)
the bias of the trial wave function which characterizes VMC calculations.  They
are a stochastic implementation of the power method for finding the dominant
eigenstate of a matrix or integral kernel. In a projector Monte Carlo method,
one uses an operator that inverts the spectrum of $\hat{H}$ to project out the
ground state of $\hat{H}$ from a given trial state.

Diffusion Monte Carlo (DMC) uses the exponential projection operator
$e^{-t(\hat{H}-E_\mathrm T)}$ with $E_\mathrm T$ a trial energy whose role will
become immediately apparent. To understand the effect of applying this operator
on a given wave function, let us consider a trial wave function $\Psi$, which we
expand on the eigenstates of $\hat{H}$, $\Psi_i$ with eigenvalues $E_i$. In the
limit of infinite time $t$, we then obtain
\begin{equation}
	\lim\limits_{t\rightarrow \infty} e^{-t(\hat{H}-E_\mathrm T)} |\Psi\rangle
	= \lim\limits_{t\rightarrow \infty} \sum_{i} e^{-t(E_i-E_\mathrm T)} |\Psi_i \rangle \langle \Psi_i | \Psi \rangle
	= \lim\limits_{t\rightarrow \infty} e^{-t(E_0-E_\mathrm T)}|\Psi_0\rangle \langle \Psi_0 | \Psi \rangle\,,
\end{equation}
where, in the last equality, we used that the coefficients in front of the
higher eigenstates decay exponentially faster than the one of the ground state.
If we adjust $E_T$ to $E_0$, the projection will yield the ground state
$\Psi_0$. Note that the starting wave function must have a non-zero overlap with
the ground-state one.

In the position representation,  this projection can be rewritten as
\begin{equation}
	\Psi(\mathbf R',t) = \int G(\mathbf R'|\mathbf R,t) \Psi(\mathbf R) \mathrm d \mathbf R\,,
\label{integral}
\end{equation}
where we introduced the coordinate Green's function defined as
\begin{equation}
G(\mathbf R'|\mathbf R,t) = \langle\mathbf R' \vert{e^{-t(\hat{H}-E_\mathrm T)}}\vert{\mathbf R}\rangle\,.
\end{equation}
This representation readily allows us to see how to translate the projection into a Markov process 
provided that we can sample the Green's function and the trial wave function.  For fermions, the
wave function is antisymmetric and cannot therefore be interpreted as a
probability distribution, a fact that we will ignore for the moment.

A further complication is that the exact form of the Green's function is not
known. Fortunately, in the limit of small-time steps $\tau$, Trotter's theorem
tells us that we are allowed to consider the potential and kinetic energy
contributions separately since
\begin{equation}
	e^{-\tau(\hat{T}+\hat{V})} = e^{-\tau\hat{T}}e^{-\tau\hat{V}} + {\cal O}(\tau^2)\,,
\end{equation}
so that
\begin{eqnarray}
       \langle \mathbf R' | e^{-\tau\hat{H}} | \mathbf R\rangle &\approx & \langle \mathbf R' | e^{-\tau\hat{T}}e^{-\tau\hat{V}} | \mathbf R\rangle 
       = \int \mathrm d\mathbf R'' \langle \mathbf R' | e^{-\tau\hat{T}} | \mathbf R''\rangle\langle \mathbf R'' | e^{-\tau\hat{V}} | \mathbf R\rangle \nonumber\\
      &=& \langle \mathbf R' | e^{-\tau\hat{T}} | \mathbf R\rangle e^{-\tau {V}(\mathbf R)}\,.
\end{eqnarray}
Therefore, we can rewrite the Green's function in the \textit{short-time approximation} as
\begin{equation}
     G(\mathbf R'|\mathbf R,\tau)= (2\pi\tau)^{-3N/2}\, \exp\left[-\frac{(\mathbf R'-\mathbf R)^2}{2 \tau}\right]\,
\exp\left[-\tau\,({V}(\mathbf R)-E_\mathrm T)\right]+{\cal O}(\tau^2)\,,
\end{equation}
where the first (stochastic) factor is the Green's function for diffusion while
the second term multiplies the distribution by a positive scalar.  
The repeated application of the short-time Green's function to obtain the
distribution at longer times (Eq.~\ref{integral}) can be interpreted as a Markov
process with the difference that the Green's function is not normalized due to
the potential term, and we therefore obtain a weighted random walk.

\begin{figure}
    \centering
    \includegraphics[width=0.4\linewidth]{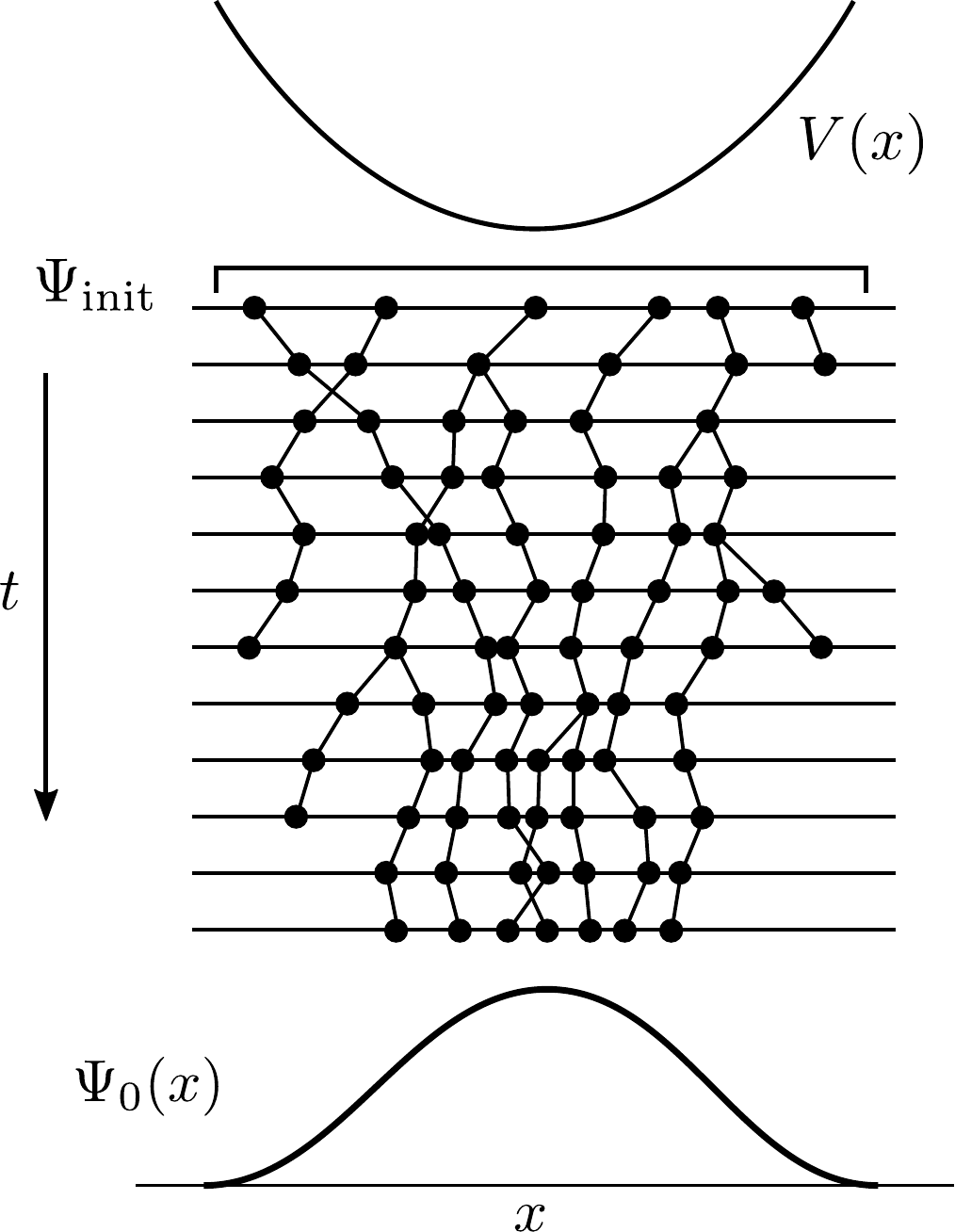}
    \caption{Schematic representation of a DMC simulation showing the evolution of the walkers in 
    a one-dimensional potential $V(x)$. The walkers are uniformly distributed at the start and converge 
    to the ground-state distribution $\Psi_0$ after a number of time steps (adapted from 
    Ref.~\cite{Foulkes2001}).}
\label{fig:dmc}
\end{figure}

The basic DMC algorithm is rather simple:
\begin{enumerate}
\itemsep0ex
\item An initial set of $M_0$ so-called walkers $\mathbf R_1,\ldots,\mathbf R_{M_0}$ is generated by
sampling the trial wave function $\Psi(\mathbf R)$ with the Metropolis algorithm as in VMC. This is
the zero-th \textit{generation} and the number of configurations is the \textit{population} of the
zero-th generation.
\item Each walker diffuses as $\mathbf R'=\mathbf R+\xi$ where $\xi$ is 
sampled from the 3$N$-dimensional Gaussian distribution $g(\xi)=(2\pi\tau)^{-3N/2}\exp\left(-\xi^2
/2\tau\right)$.
\item For each walker, we compute the factor
\begin{equation}
p=\exp\left[-\tau({V}(\mathbf R)-E_{\mathrm T})\right]\,.
\end{equation}
and perform the so-called branching step, namely, we branch the walker by treating $p$ as the probability to
survive at the next step: if $p<1$, the walker survives with probability $p$ while, if $p>1$, the walker
continues and new walkers with the same coordinates are created with probability $p-1$. This is achieved
by creating a number of copies of the current walker equal to the integer part of  $p+\eta$ where
$\eta$ is a random number between (0,1).
The branching step causes walkers to live in regions 
with a low potential ${V} < E_\mathrm T$ and die in regions with high $V$. 
\item We adjust $E_\mathrm T$ so that the overall population fluctuates around
the target value $M_0$.
\end{enumerate}
Steps 2-4 are repeated until a stationary distribution is obtained and the desired properties are 
converged within a given statistical accuracy. A schematic representation of the evolution for a simple
one-dimensional problem is shown in Fig.~\ref{fig:dmc}.
Since the short-time expression of the Green's function is only valid in the limit of $\tau$ approaching 
zero, in practice, DMC calculations must be performed for different values of $\tau$ and the results
extrapolated for $\tau$ which goes to zero.

The direct sampling of this Green's function proves however to be highly inefficient and unstable since 
the potential can vary significantly from configuration to configuration or also be unbounded like the Coulomb 
potential. For example, the electron-nucleus potential diverges to minus infinity as the two particles approach 
each other, and the branching factor will give rise to an unlimited number of walkers. Even if the potential is 
bounded, the approach becomes inefficient with increasing size of the system since the branching factor also 
grows with the number of particles.  These difficulties can be overcome by using \textit{importance sampling}, where 
the trial wave function, $\Psi$, is used to guide the random walk. Starting from Eq.~\ref{integral},
we multiply each side by $\Psi(\mathbf R')$ and define the probability distribution $f(\mathbf R,t)=\Psi(\mathbf R,t)\Psi(\mathbf R)$ 
which satisfies
\begin{equation}
f(\mathbf R',t) = \int \tilde G(\mathbf R'|\mathbf R,t) \Psi(\mathbf R)^2 \mathrm d \mathbf R\,,
\end{equation}
where the importance sampled Green's function is given by
\begin{equation}
\tilde{G}(\mathbf R'|\mathbf R,t) = \Psi(\mathbf R')\langle{\mathbf R}'
\vert{e^{-t(\hat{H}-E_\mathrm T)}}\vert{\mathbf R}\rangle/\Psi({\mathbf R})\,.
\end{equation}
In the limit of long times, this distribution $f(\mathbf R,t)$ approaches $\Psi_0(\mathbf R)\Psi(\mathbf R)$. 

Assuming for the moment that $\Psi(\mathbf R')/\Psi(\mathbf R) > 0$, the importance sampled Green's function 
in the short-time approximation becomes
\begin{align}
	\tilde G(\mathbf R'|\mathbf R, \tau) \approx (2\pi\tau)^{-\frac{3}{2}N} 
	\exp \left[ -\frac{(\mathbf R' - \mathbf R - \tau \mathbf V(\mathbf R))^2}{2\tau} \right]
	\exp \left[ -\tau(E_\mathrm L(\mathbf R) - E_\mathrm T ) \right]\,,
\end{align}
where one has assumed that the drift-velocity $\mathbf V(\mathbf R) = \nabla
\Psi(\mathbf R)/\Psi(\mathbf R)$ and the local energy (Eq.~\ref{eq:rho}) are constant in the step
from $\mathbf R$ to $\mathbf R'$. There are two important new features of
$\tilde{G}$. First, the quantum velocity $\mathbf V(\mathbf R)$ pushes the
walkers to regions where $\Psi(\mathbf R)$ is large. In addition, the local
energy $E_\mathrm L$ instead of the potential appears in the branching factor.
Since the local energy becomes constant and equal to the eigenvalue as the trial
wave function approaches the exact eigenstate, we expect that, for a good trial
wave function,  the fluctuations in the branching factor will be significantly
smaller. In particular, imposing the cusp conditions on the wave function will
remove the instabilities coming from the singular Coulomb potential. The simple
DMC algorithm can be easily modified by sampling the square of the trial wave
function in a VMC calculation (step 1), drifting before diffusing the walkers
(step 2), and employing the exponential of the local energy as branching factor
(step 3). Several important modifications to this bare-bone algorithm can and
should be introduced to reduce the time-step error, which are described in
detail along with further improvements in Ref.~\cite{Umrigar1993}.

\begin{figure}[t]
	\centering
	\includegraphics[width=0.5\linewidth]{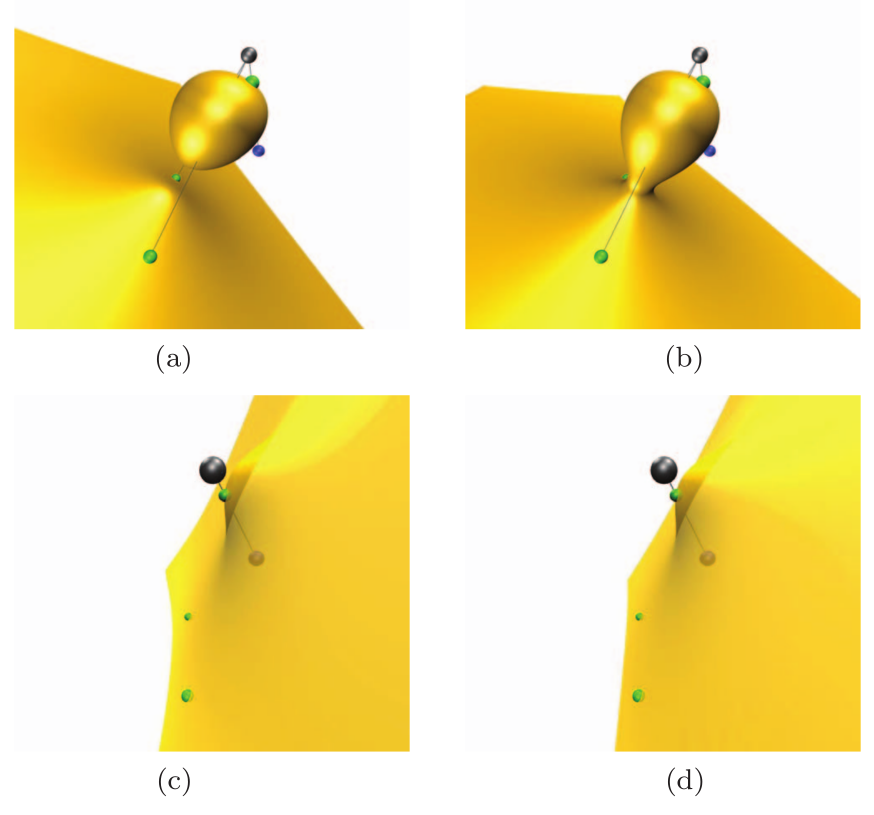}
	\caption{A 3-dimensional slice of the nodal surface of the N (top) and P (bottom) atoms 
        obtained as a scan of the wave function moving one electron and keeping the others at
        snapshot positions (green/blue spheres). The black spheres mark the positions of the nucleus.
        (a, c) Nearly exact nodes and (b, d) Hartree-Fock nodes. 
        Adapted from Ref.~\cite{rasch_fixednode_2014}.}
	\label{fig:nodes}
\end{figure}
Up to this point, we have assumed that the wave function does not change sign.
This is true for the ground state of a bosonic system, whose wave function can
be in principle projected exactly in a DMC simulation.  For fermions, however, a
move of a walker can lead to a change of sign due to the antisymmetry of the
wave function. While it is possible to work with weights that carry a sign, the
stochastic realization of such a straightforward approach is not stable since
the separate evolution of the populations of positive and negative walkers will
lead to the same bosonic solution, and the fermionic signal will be
exponentially lost in the noise. This is known as the \textit{fermionic sign
problem}. To circumvent this problem, we can simply forbid moves in which the
sign of the trial wave function changes and the walker crosses the nodes which
are defined as (3\textit{N}$-$1)-dimensional surface where the trial wave
function is zero. Imposing the nodal constraint can be achieved either by
deleting the walkers which attempt to cross the nodes or by using the short-time
importance sampled Green's function, where walkers do not cross the nodes in the
limit of zero time step. This procedure is known as the \textit{fixed-node approximation}. Forbidding node crossing is equivalent to finding the exact
solution with the boundary condition of having the same nodes as the trial wave
function.  The Schr\"odinger equation is therefore solved exactly inside the
nodal regions but not at the nodes where the solution will have a discontinuity
of the derivatives.  The fixed-node solution will be exact only if the nodes of
the trial wave function are exact. In general, the fixed-node energy will be an
upper bound to the exact energy. A cut through the nodal surface of the N and P
atoms for a simple Hartree-Fock and a highly-accurate wave function
(Fig.~\ref{fig:nodes}) reveals that considerable differences are possible which
are atom dependent and directly translate in a larger size of the fixed-node
error for the N atom when a mono-determinantal wave function is used~\cite{rasch_fixednode_2014}.

The fixed-node DMC algorithm can be also used to study excited states. There is
no particular difficulty in applying DMC to the lowest state of a given symmetry
by simply employing a trial wave function of the proper spatial and spin
symmetry\footnote{More precisely, the DMC energy is variational if the trial function transforms according to a one-dimensional irreducible representation of the symmetry group of the Hamiltonian~\cite{Foulkes1999}.}. For excited states which are
energetically not the lowest in their symmetry, all that we know is that
fixed-node DMC will give the exact solution if we employ a trial wave function
with the exact nodes~\cite{Foulkes1999}. However, there is no variational
principle and one may expect a stronger dependence of the result on the choice
of the wave function, which is now not only used to overcome the fermion-sign
problem but also to select the state of interest. In our experience, unless we
intentionally generate a wave function with a large overlap with the
ground-state one, we do not suffer from lack of variationality in the
excited-state calculation.  In fact, the use of simplistic wave functions (e.g.\
HOMO-LUMO Hartree-Fock, configuration-interaction singles, non-reoptimized
truncated complete-active-space expansions) has been shown to generally lead to
an overestimation of the excitation energy also in DMC, especially when the
excited state has a strong multi-determinant
character~\cite{schautz_excitations_2004}. Consequently, while DMC cannot cure
the shortcomings of a poor wave function, such a choice will likely yield
smaller fixed-node errors in the ground than the excited state and, ultimately,
an overestimation of the DMC excitation energy.

\section{Wave functions and their optimization}
\label{sec:wf-optimization}

The key quantity which determines the quality of a VMC and a fixed-node DMC
calculation is the trial wave function.  The choice of the functional form of
the wave function and its optimization within VMC are key steps in a QMC
calculation as they are crucial elements to obtain accurate results already at
the VMC level and to reduce the fermionic-sign error in a subsequent DMC
calculation.

Most QMC studies of electronic systems have employed trial wave functions of the
so-called Jastrow-Slater form, that is, the product of a sum of determinants of
single-particle orbitals and a Jastrow correlation factor as
\begin{equation}
\Psi = \mathcal{J} \sum_k c_k D_k\,,
\label{JS}
\end{equation}
where $D_k$ are Slater determinants of single-particle orbitals and the Jastrow
correlation function is a positive function of the interparticle distances,
which explicitly depends on the electron-electron separations. The Jastrow
factor plays an important role as it is used to impose the Kato cusp conditions
and to cancel the divergences in the potential at the inter-particle coalescence
points. This leads to a smoother behavior of the local energy and therefore more
accurate and efficient VMC as well as DMC calculations thanks to the smaller
time-step errors and reduced fluctuations in the branching factor.

Moreover, the Jastrow factor introduces important correlations beyond the short
elec\-tron-electron distances~\cite{Prendergast2001} and QMC wave functions
enjoy therefore a more compact determinantal expansion than conventional quantum
chemical methods. Even though the positive Jastrow function does not directly
alter the nodal structure of the wave function which is solely determined by the
antisymmetric part, the optimal determinantal component in a QMC wave function
will be different than the one obtained for instance in a multi-configuration
self-consistent-field calculation (MCSCF) in the absence of the Jastrow factor. 
Upon optimization of the QMC wave function, the determinantal component will
change and it is often possible to obtain converged energy differences in VMC
and DMC with relatively short determinantal expansions in a chosen active space.
Furthermore, thanks to the presence of the Jastrow factor, QMC results are
generally less dependent on the basis set. For instance, excitations and
excited-state gradients show a faster convergence with basis set than
multiconfigurational approaches, and an augmented double basis set with polarization
functions
is often sufficient in both VMC and DMC for the description of excited-state
properties~\cite{valsson_photoisomerization_2010, Zulfikri2016, blunt_excited-state_2019}.

Important requirement for the optimization of the many parameters in a QMC wave
function is the ability to efficiently evaluate the derivatives of the wave
function and the action of the Hamiltonian on these derivatives during a QMC
run.  In general, this is central to the computation of low-variance estimators
of derivatives of the total energy as, for instance, the derivatives with
respect to the nuclear coordinates (i.e.\ interatomic forces).  Computing these
derivatives at low cost is therefore crucial to enable higher accuracy as well
as to extend the application of QMC to larger systems and a broader range of
molecular properties. Automatic differentiation was successfully applied for the
computation of analytical derivatives~\cite{Sorella2010} but the application to
large computer codes is not straightforward and the memory requirements might
become prohibitive.

Recently, an efficient and simple analytical formulation has been developed to compute a complete set of 
derivatives of the wave function and of the local energy with the same scaling per Monte Carlo step as 
computing the energy alone both for single- and multi-determinant wave functions~\cite{Filippi2016,assaraf_optimizing_2017}. 
This formulation 
relies on the straightforward manipulation of matrices evaluated on the occupied and virtual orbitals and can be 
very simply illustrated in the case of a single determinant in the absence of a Jastrow factor:
\begin{equation}
D = \det(A) = |\phi_1 \phi_2 \ldots \phi_N|\,,
\end{equation}
where $A$ is a Slater matrix defined in terms of the $N$ occupied orbitals, $\phi_i$, as $A_{ij}=\phi_j(\mathbf r_i)$. 
For this wave function, it is not difficult to show that the action of a one-body operator 
$\hat{O}=O(\mathbf r_1)+\ldots+O(\mathbf r_N)$ on the determinant can be written as the trace between the inverse 
$A$ matrix and an appropriate matrix $B$,
\begin{equation}
  \frac{\hat OD}{D} = \tr(A^{-1}B)\,,
  \label{trace}
\end{equation}
where $B$ is obtained by applying the operator $O(\mathbf r)$ to the elements of $A$ as 
\begin{equation}
  B_{ij}=(O\phi)_j(\mathbf r_i)\,.
\end{equation}
For instance, if we consider the kinetic operator, we obtain
\begin{eqnarray}
   \hat{T} \det(A)= -\frac{1}{2} \sum_i \Delta_i \det(A) 
   = -\frac{1}{2} \sum_i\;\left[\sum_j \Delta \phi_j({\mathbf r}_i)\, (A^{-1})_{ji}\det(A)\right]\,,
\end{eqnarray}
which can be rewritten as
\begin{eqnarray}
  \frac{\hat{T}\det(A)}{\det(A)} = \sum_i\sum_j B_{ij}^\mathrm{kin} \,(A^{-1})_{ji} = \tr(A^{-1} B^\mathrm{kin})\,,
\end{eqnarray}
where $B^\mathrm{kin}_{ij}=-\frac{1}{2} \Delta_i A_{ij}=-\frac{1}{2}\Delta \phi_j({\mathbf r}_i)$. 
It is possible to show that an equivalent trace expression holds also in the presence of the Jastrow factor but with a 
$B$ matrix which depends not only on the orbitals but also on the Jastrow factor.

\begin{figure}
    \includegraphics[width=0.35\textwidth,angle=-90]{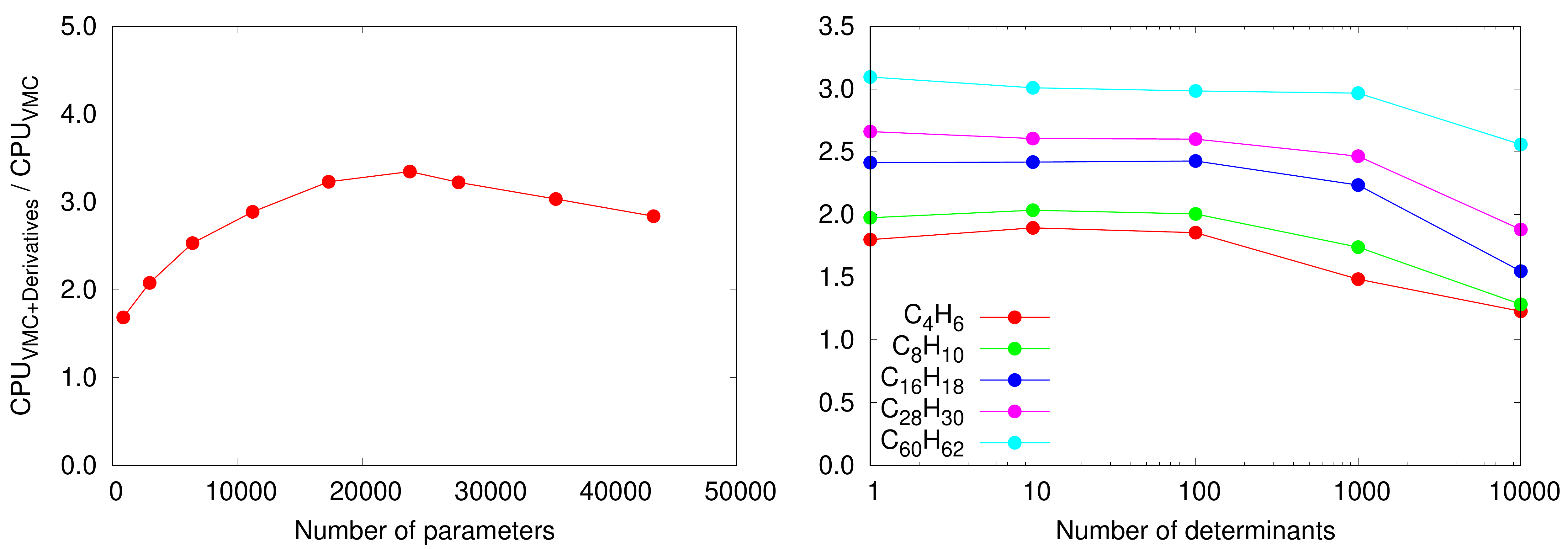}
    \caption{Cost per Monte Carlo step of computing a complete set of derivatives of the 
    wave function and local energy relative to a VMC run where only the energy is calculated. 
    Left: increasing number of variational parameters for the series $\text{C}_n\text{H}_{n+2}$ 
    with $n$ between 4 and 44 (reproduced from Ref.~\cite{Filippi2016}). Right: increasing number 
    of determinants in the Jastrow-Slater wave function for $\text{C}_n\text{H}_{n+2}$ with $n$ 
    between 4 and 60 (reproduced from Ref.~\cite{assaraf_optimizing_2017}).}
\label{fig:derivatives}
\end{figure}

The compact trace expression of a local quantity (Eq.~\ref{trace}) offers the advantage that its derivative 
with respect to a parameter $\mu$ can be straightforwardly written as
\begin{eqnarray} 
  \partial_\mu \frac{\hat OD}{D} = \tr({A}^{-1} \partial_\mu B-X\partial_\mu A)\,,
\end{eqnarray}
where $\partial_\mu A$ and $\partial_\mu B$ are the matrices of the derivatives of the elements of $A$ and $B$, respectively, 
and the matrix $X$ is defined as
\begin{eqnarray}
  X=A^{-1}{B}A^{-1}\,.
\end{eqnarray}
This can easily be derived by using $\partial_\mu(A^{-1})=-A^{-1}\partial_\mu
A\,A^{-1}$ and the cyclic property of the trace.  Therefore, if one computes and
stores the matrix $X$, it is then possible to evaluate derivatives at the
cost of calculating a trace, namely, $O(N^2)$.  Consequently, this procedure
enables for instance the efficient calculation of the $O(N)$ derivatives of the
local energy ($\hat O = \hat H$) with respect to the nuclear coordinates, reducing the scaling of
computing the interatomic forces per Monte Carlo step to the one of the energy,
namely, $O(N^3)$. The same scaling is also obtained for the optimization of the
orbital parameters as shown in Fig.~\ref{fig:derivatives} (left panel) and further 
discussed in Ref.~\cite{Filippi2016}.  This simple
formulation and its advantages in the calculation of energy derivatives can be extended
to multi-determinant wave functions to achieve a cost in the computation of a
set of derivatives proportional to the one of evaluating the
energy alone~\cite{assaraf_optimizing_2017} as illustrated in Fig.~\ref{fig:derivatives} 
(right panel) for the interatomic derivatives.

With all the derivatives of the wave function and the corresponding local
quantities at hand, the next step is to use them for the optimization of the
wave function. The use of wave functions with a large number of parameters
requires efficient algorithms and the two most commonly used approaches, the
linear method and the stochastic reconfiguration method, are discussed in the
following.  We will begin with the simpler case of the optimization in the
ground state (or an excited state which is energetically the lowest for a given
symmetry).

\subsubsection{Stochastic Reconfiguration Method}
\label{sec:sr}

In the stochastic reconfiguration (SR) method~\cite{Sorella2001,Sorella2004}, one starts from a given wave function, 
$\Psi$, and obtains an improved state by applying the operator $(1-\tau \hat{H})$, namely, a first-order expansion 
of the operator $e^{-\tau\hat{H}}$ used in DMC. The new state is then projected in the space spanned by the 
current wave function and its derivatives, $\{\Psi_i\}=\{\Psi,\partial_i\Psi\}$ as
\begin{equation}
\sum_{j=0}^{N_p} \delta p_j |\Psi_j \rangle = \hat{P}^\text{SR} (1-\tau \hat{H}) | \Psi \rangle \,,
\end{equation}
where $N_p$ is the number of wave function parameters and $\Psi_0=\Psi$. By taking the internal product with $\langle \Psi_i|$ and
eliminating the scaling $\delta p_0$ through the $i=0$ equation, one derives a set of equations for $i=1,\ldots,N_p$, 
which can be written in matrix notation as
\begin{equation}
\bar{\mathbf S}\,\delta\mathbf p=-\frac{\tau}{2}\mathbf g\,,
\label{eq:sr}
\end{equation}
where $\mathbf g$ is the gradient of the energy with respect to the parameters,
\begin{equation}
g_i=2\left[\frac{\langle \Psi_i | \hat{H} | \Psi \rangle}{\langle \Psi | \Psi \rangle} -E_\mathrm v\frac{\langle \Psi_i | \Psi \rangle}{\langle \Psi | \Psi \rangle}\right] \equiv \partial_i E_\mathrm v\,,
\label{gradient}
\end{equation}
and $\bar{\mathbf S}$ is related to the overlap matrix between the derivatives, $\mathbf S$, as
\begin{equation}
\bar{S}_{ij}=\frac{\langle \Psi_i | \Psi_j \rangle}{\langle \Psi|\Psi\rangle}-\frac{\langle \Psi_i|\Psi\rangle}{\langle \Psi|\Psi\rangle}\frac{\langle \Psi|\Psi_j\rangle}{\langle \Psi|\Psi\rangle}\equiv S_{ij}-S_{i0}S_{0j}
\label{eq:Smat}
\end{equation}
With an appropriate choice of $\tau$, a new set of parameters can be determined as
$p^\prime_i=p_i + \delta p_i$ and the procedure iterated until convergence.  Therefore, the SR method is
like a Newton approach where one follows the downhill gradient of the energy, using however the matrix 
$\bar{\mathbf S}$ instead of the Hessian of the energy.  Even though the method can display a slow convergence 
since $\tau$ scales like the inverse of the energy range spanned by the wave function derivatives~\cite{schautz_optimized_2004}, it 
was recently employed to successfully optimize very large numbers of parameters~\cite{assaraf_optimizing_2017,dash_perturbatively_2018}.
We will come back to this point when discussing the linear method below.

In a VMC run of SR optimization, one needs to compute the gradient and the overlap matrix ${\mathbf S}$ by
sampling the distribution $\rho$ given by the square of the current wave function (Eq.~\ref{eq:rho}) as
\begin{eqnarray}
S_{ij}&=&\frac{\langle \Psi_i | \Psi_j \rangle}{\langle \Psi | \Psi \rangle} 
=\frac{\int \Psi_i(\mathbf R) \Psi_j(\mathbf R)\mathrm d\mathbf R}{\int \Psi(\mathbf R)^2\mathrm d\mathbf R}\nonumber\\
&=&\int \frac{\Psi_i (\mathbf R)}{\Psi (\mathbf R)}
      \frac{\Psi_j (\mathbf R)}{\Psi (\mathbf R)}\rho(\mathbf R)\mathrm d\mathbf R
\approx\frac{1}{M} \sum_k^{M} 
        \frac{\Psi_i (\mathbf R_k)}{\Psi (\mathbf R_k)}
        \frac{\Psi_j (\mathbf R_k)}{\Psi (\mathbf R_k)}\,.
\end{eqnarray}
For large number of parameters, not only the storage of this matrix becomes problematic but also its calculation
whose cost scales as $\mathcal O(M N_p^2)$. However, if we use a conjugate gradient method to solve the 
linear equations (\ref{eq:sr}), we only need to repeatedly evaluate ${\mathbf S}$ acting on a trial 
vector of parameter variations as
\begin{equation}
\sum_{j=1}^{N_p}{S}_{ij}\delta p_j 
=       \frac{1}{M} \sum_k^{M} 
        \frac{\Psi_i (\mathbf R_k)}{\Psi (\mathbf R_k)} \sum_{j=1}^{N_p}
        \frac{\Psi_j (\mathbf R_k)}{\Psi (\mathbf R_k)}\,\delta p_j\,,
\end{equation}
where the order of the sums in the last expression has been swapped~\cite{Neuscamman2012}.
Therefore, if we compute and store the $M\times N_p$ matrix of the ratios $\Psi_i
(\mathbf R_k)/\Psi (\mathbf R_k)$ during the Monte Carlo run, we can reduce the
memory requirements by exploiting the intrinsic parallelism of Monte Carlo
simulations: we can employ a small $M$ per core and increase instead the number
of cores to obtain the desired statistical accuracy. The computational cost of
solving the SR equations is also reduced to $\mathcal{O}(N_\text{CG}M N_p)$,
where $N_\text{CG}$ is the number of conjugate gradient steps, which we have
found to be several orders of magnitude smaller than the number of parameters in
recent optimization of large determinantal expansions~\cite{assaraf_optimizing_2017,dash_perturbatively_2018}.

\subsubsection{Linear Method}
\label{sec:lm}

The linear optimization method is related to the so-called super configuration interaction (super-CI) approach used in quantum 
chemistry to optimize the orbital parameters in a multi-determinant wave function.  The starting point is the 
normalized wave function, 
\begin{equation}
|\bar \Psi\rangle = \frac{1}{\sqrt{\langle \Psi| \Psi\rangle}}|\Psi\rangle\,,
\end{equation}
which we expand to first order in the parameter variations around the current values 
in the basis of the current wave function and its derivatives, $\{\bar\Psi_i\}=\{\bar \Psi,\partial_i \bar \Psi\}$ with $\bar \Psi_0=\bar \Psi$. 
The important advantage of working with the ``barred'' functions is that they are orthogonal to the current wave function
since 
\begin{equation}
|\bar \Psi_i \rangle = \frac{1}{\sqrt{\langle \Psi| \Psi\rangle}}\big(|\Psi_i \rangle - \frac{\langle \Psi | \Psi_i \rangle}{\langle \Psi | \Psi\rangle} |\Psi \rangle\big)\,,
\label{eq:semi-ortho}
\end{equation}
which has been found to yield better (non-linear) parameter variations and a more robust optimization than simply using the derivatives of the wave function.

The change of the parameters $\delta \mathbf p$ is then determined by minimizing the expectation value of the Hamiltonian on the linearized wave function in the basis $\{\bar \Psi_i\}$, which leads to the generalized eigenvalue equations:
\begin{equation}
\sum_{j=0}^{N_p} \bar{H}_{ij}\delta p_j=E_\text{lin}\sum_{j=0}^{N_p} \bar{S}_{ij}\delta p_j\,,
\end{equation}
where $\bar{H}_{ij} = \langle \bar \Psi_i | \hat H | \bar \Psi_j \rangle$ and
$\bar{S}_{ij} = \langle \bar \Psi_i | \bar \Psi_j \rangle$. We note that the
overlap $\bar{\mathbf S}$ is equivalent to the expression introduced above in
the SR scheme (Eq.~\ref{eq:Smat}). A new set of parameters can be generated as
$p_i^\prime=p_i+\delta p_i/\delta p_0$ and the algorithm iterated until
convergence. Importantly, in a Monte Carlo run, the matrix $\bar{\mathbf H}$
will not be symmetric for a finite sample and a non-obvious finding is that the
method greatly benefits from reduced fluctuations if one does \textit{not}
symmetrize the Hamiltonian matrix, as originally shown by Nightingale and Melik-Alaverdian
for the optimization of the linear parameters~\cite{Nightingale2001}.  Other
important modifications can be introduced to further stabilize the approach and
improve the convergence as discussed in Ref.~\cite{Toulouse2007a}.

It is simple to recognize that, at convergence, the linear method leads to an
optimal energy if we express explicitly the secular equations above in matrix
form as
\begin{equation}
\begin{pmatrix}
E_\mathrm v  & \frac{1}{2}g^\text T \\
\frac{1}{2}g & \bar{\mathbf H}
\end{pmatrix}
\begin{pmatrix}
\delta p_0 \\
\delta \mathbf p
\end{pmatrix} = E_\text{lin}
\begin{pmatrix}
1 & 0 \\
0 & \bar{\mathbf S}
\end{pmatrix}
\begin{pmatrix}
\delta p_0 \\
\delta \mathbf p
\end{pmatrix}\,,
\label{eq:lm-eigenvalue}
\end{equation}
where we have used that $\bar H_{00}= \langle \bar \Psi | \hat H | \bar \Psi \rangle$ is the current energy
and the elements of the first column and row, $\bar H_{i0}=\langle \bar \Psi_i | \hat H | \bar \Psi \rangle$ and 
$\bar H_{0i}=\langle \bar \Psi | \hat H | \bar \Psi_i \rangle$, respectively, are both mathematically equal to the components of
the energy gradient (Eq.~\ref{gradient}). Therefore, when the wave function parameters are optimal, the variations with 
respect to the current wave function will no longer couple to it ($\delta p_i=0$) and the $\bar H_{i0}$ and $\bar H_{0i}$ 
elements must therefore become zero.  This directly implies that the gradient of the energy with respect to the parameters 
is identically zero.

To further understand how the linear method is related to other optimization schemes, one can recast its equations as a Newton method~\cite{Toulouse2008}:
\begin{equation}
(\mathbf A +\alpha \bar{\mathbf S})\delta\mathbf p=-\frac{1}{2}\mathbf g\,,
\end{equation}
where $\mathbf A= \bar{\mathbf H} -E_\mathrm v \bar{\mathbf S}$  and $\alpha=E_\mathrm v-E_\text{lin}>0$.
Therefore, the parameters are varied along the downhill gradient of the energy
with the use of an approximate Hessian $\mathbf A$ level-shifted by the positive definite matrix $\alpha
\bar{\mathbf S}$. The presence of the latter renders
the optimization more stable and effective than the actual Newton method even when
the exact Hessian matrix is used.
While the linear method is in principle significantly
more efficient than for instance the SR approach, we find that its stochastic realization
suffers from large fluctuations in the elements of $\bar{\mathbf H}$ when one optimizes the 
orbital parameters or the linear coefficients of particularly extended multi-determinant wave functions (where the
derivatives are very different from the actual wave function used in the sampling). As 
a result, the optimization requires long VMC runs to achieve reliable variations in the
parameters or a large shift added to the diagonal elements of $\bar{\mathbf H}$  (except $\bar{H}_{00}$)~\cite{Toulouse2007a} to stabilize the procedure. In these cases,
we find that the SR scheme, which only makes use of the $\bar{\mathbf S}$ matrix, is more robust and 
efficient since it allows less strict requirements on the error bars.

Finally, we note that, as in the SR scheme, it is possible to avoid to explicitly construct the full
matrices $\bar{\mathbf H}$ and $\bar{\mathbf S}$: one stores the local
quantities $\Psi_i(\mathbf R)/\Psi(\mathbf R)$ and $\hat{H}\Psi_i(\mathbf
R)/\Psi(\mathbf R)$ in the Monte Carlo run and uses for instance a generalized
Davidson algorithm to find the eigenvectors where only matrix-vector products
with trial vectors are evaluated, significantly reducing the computational and
memory requirements~\cite{Neuscamman2012}. 

\section{Excited States}
\label{sec:excited}

For excited states of a different symmetry than the ground state, one can construct a trial wave function of the
desired space and spin symmetry (with an appropriate choice of the determinantal component) and apply either the SR 
or the linear method to minimize the energy in VMC, subsequently refining the calculation in DMC. For excited states 
which are energetically not the lowest in their symmetry class, one can instead follow different routes as in other 
quantum chemistry methods to find an accurate excited-state wave function. We will begin to describe the possibilities 
within energy minimization and then consider optimization schemes targeting the variance of the energy which has a minimum for 
each eigenstate of the Hamiltonian.

\subsection{Energy-based methods}
\label{sec:enemin}

While the linear method is generally employed for ground-state wave function optimizations, it is in fact 
possible to use it in a state-specific manner for the optimization of excited states~\cite{zimmerman_excited_2009}.
One can target a higher-energy state and linearize the problem with respect to the chosen state. Solving a generalized 
eigenvalue problem as in Eq.~\ref{eq:lm-eigenvalue} will yield lower energy roots as well as the state of interest. The
resulting wave function will be only approximately orthogonal to the lower ones since orthogonality is only imposed 
in the basis of the variations of the optimal target wave function with respect to the parameters. Furthermore, since 
following such a higher root leads to the optimization of a saddle point in parameter 
space, the procedure may exhibit convergence problems so that the parameters do not converge to the desired
state. One may also observe flipping of the roots: As the optimization proceeds the optimized excited target state 
can obtain a lower eigenvalue than the unoptimized ground state. Such problems will be particularly severe in case of close degeneracy as in proximity of conical intersection regions. 

A different route to optimize multiple states of the same symmetry lies in the generalization of state-average (SA)
approaches to QMC~\cite{filippi_absorption_2009}. We start from a set of Jastrow-Slater wave functions for the multiple 
states that are constructed as linear combinations of determinants multiplied by a Jastrow factor as
\begin{equation}
\Psi^I = \mathcal J \sum_{k} c_k^ID_k\,,
\end{equation}
where the index $I$ labels the states. The wave functions of the different states are therefore characterized
by different linear coefficients $c_i^I$ but share a common set of orbitals and the Jastrow factor 
$\mathcal J$. 

The optimal linear coefficients $c_i^I$ can be easily determined through the solution of the generalized eigenvalue equations
\begin{equation}
    \sum_{j} H_{ij}c_j^I = E_I \sum_{j} S_{ij} c_j^I\,,
\label{eq:state-average}
\end{equation}
where the matrix elements are here given by 
\begin{equation}
    H_{ij}=\langle \mathcal J D_i |\hat{H}| \mathcal J D_j\rangle \ \ \ \text{and}\ \ \ S_{ij}=\langle \mathcal J D_i | \mathcal J D_j\rangle\,,
\end{equation}
and are computed in a VMC run, where we do not symmetrize the Hamiltonian matrix
for finite Monte Carlo sampling to reduce the fluctuations of the parameters as
discussed above for the general linear method.  After diagonalization of
Eq.~\ref{eq:state-average}, the optimal linear coefficients are obtained and, at
the same time, orthogonality between the individual states is automatically
enforced.

To obtain a robust estimate of the linear coefficients of multiple states, it is
important that the distribution sampled to evaluate $H_{ij}$ and $S_{ij}$ has a
large overlap with all states of interest (and all lower lying states). A
suitable guiding wave function can for instance be constructed as
\begin{equation}
   \Psi_g=\sqrt{\sum_I |\Psi_I|^2}\,,
\label{eq:psig}
\end{equation}
and the distribution $\rho_g$ in the VMC run defined as the square of this guiding function. 
The matrix elements $S_{ij}$ (and, similarly, $H_{ij}$) are then evaluated in the Monte Carlo run as
\begin{eqnarray}
    \frac{S_{ij}}{\langle\Psi_g|\Psi_g\rangle}
    &=&\frac{\int \mathcal J D_i(\mathbf R) \mathcal J D_j(\mathbf R)\mathrm d\mathbf R}{\int \Psi_g^2(\mathbf R)\mathrm d\mathbf R}
    =\int \frac{\mathcal J D_i(\mathbf R)}{\Psi_g(\mathbf R)} \frac{\mathcal J D_j(\mathbf R)}{\Psi_g(\mathbf R)}\rho_g(\mathbf R)\mathrm d\mathbf R \nonumber\\
    &\approx& \frac{1}{M} \sum_k^M \frac{\mathcal J D_i(\mathbf R_k)}{\Psi_g(\mathbf R_k)} \frac{\mathcal J D_j(\mathbf R_k)}{\Psi_g(\mathbf R_k)} \,.
\end{eqnarray}
We note that we can introduce the 
denominator $\langle\Psi_g|\Psi_g\rangle$ if we simply divide by it both sides of Eq.~\ref{eq:state-average}.  

As done in state-average multi-configurational approaches to obtain a balanced description of the states of interest, one can optimize
the non-linear parameters of the orbitals and the Jastrow factor by minimizing the state-average energy
\begin{equation}
    E^\text{SA} = \sum_I w_I \frac{\langle \Psi^I|\hat{H}|\Psi^I \rangle}{\langle \Psi^I | \Psi^I \rangle}\,,
\label{eq:saE}
\end{equation}
with the weights of the states $w_I$ kept fixed and $\sum_I w_I=1$. The gradient of the SA energy can be rewritten as
\begin{equation}
    g^\text{SA}_i = \sum_I w_I {\langle \bar{\Psi}^I_i |\hat{H}|\Psi^I \rangle}\,,
\end{equation}
where, similarly to Eq.~\ref{eq:semi-ortho}, we have introduced for each state the variations, ${|\Psi_i^I\rangle=|\partial_i\Psi^I\rangle}$, 
and the corresponding ``barred'' functions orthogonal to the current state $| \Psi^I \rangle$:
\begin{equation}
    |\bar \Psi_i^I \rangle= \frac{1}{\sqrt{\langle \Psi^I | \Psi^I \rangle}}\big(|\Psi_i^I\rangle - \frac{\langle \Psi^I|\Psi^I_i\rangle}{\langle \Psi^I|\Psi^I\rangle}|\Psi^I\rangle\big)\,.
\end{equation}

The variations in the parameters can be obtained as the lowest-energy solution of the generalized eigenvalue equation in analogy to the 
linear method for the ground state,
\begin{equation}
\begin{pmatrix}
E^\text{SA}            & \frac{1}{2}(g^\text{SA})^\text T \\
\frac{1}{2}g^\text{SA} & \bar{\mathbf H}^\text{SA}
\end{pmatrix}
\begin{pmatrix}
\delta p_0 \\
\delta \mathbf p
\end{pmatrix} = E
\begin{pmatrix}
1 & 0 \\
0 & \bar{\mathbf S}^\text{SA}
\end{pmatrix}
\begin{pmatrix}
\delta p_0 \\
\delta \mathbf p
\end{pmatrix}\,.
\label{eq:eqsa}
\end{equation}
The state-average matrix elements are defined as
\begin{equation}
    \bar H_{ij}^\text{SA} = \sum_I w_I \frac{\langle \bar \Psi^I_i | \hat{H} | \bar \Psi^I_j \rangle}{\langle \bar \Psi^I |\bar \Psi^I \rangle}\,,
\end{equation}
and an analogous expression for $\bar S_{ij}^\text{SA}$ is introduced. To compute
these matrix elements in VMC, we perform a single run sampling the square of a guiding wave
function $\Psi_g$ (Eq.~\ref{eq:psig}) and compute the numerators and
denominators in the matrix expressions for all relevant states. We note that the
state-average equations (Eq.~\ref{eq:eqsa}) are not obtained by minimizing a linearized expression of the SA
energy (Eq.~\ref{eq:saE}) but are simply inspired by the single-state case.
However, since the first row and column in Eq.~\ref{eq:eqsa} are given by
the gradient of the SA energy, at convergence, the optimal parameters minimize
the SA energy. We find that the use of these state-average Hamiltonian and
overlap matrices leads to a similar convergence behavior as the linear method
for a single state.

Following this procedure, the algorithm alternates between the minimization of
the linear and the non-linear parameters until convergence is reached. 
The obtained energy $E_\text{SA}$ is stationary with respect to variations of all
parameters while the energies of the individual states, $E_I$, are only stationary
with respect to the linear but not the orbital and Jastrow parameters.  If the
ground state and the target excited state should be described by very different
orbitals, a state-specific approach may yield more accurate energies.

\subsection{Time-dependent linear-response VMC}
\label{sec:tdvmc}

A very different approach to the computation of multiple excited states is a VMC
formulation of linear-response theory~\cite{mussard_chapter_2018}. Given a
starting wave function $\Psi$ with optimal parameters $\mathbf p_0$, a
time-dependent perturbation $\hat V(t)$ is introduced in the Hamiltonian $\hat
H$ with the coupling constant $\gamma$ as
\begin{equation}
    \hat H(t) = \hat H + \gamma \hat V(t)\,,
\end{equation}
so that the ground-state wave function itself becomes time-dependent as the variational 
parameters $\mathbf p(t)$ evolve in time. It is convenient to work with a wave function 
subject to an intermediate normalization,
\begin{equation}
    |\bar \Psi(t)\rangle = \frac{|\Psi(t)\rangle}{\left<\Psi_0 | \Psi(t) \right>}\,,
\label{eq:barwf}
\end{equation}
where the starting wave function $\Psi_0\equiv\Psi$ is taken to be normalized. This choice 
leads to wave function variations to first and second order that are orthogonal to the 
current optimal wave function $\Psi_0$.

At each time $t$, one can apply the Dirac-Frenkel variational principle to obtain the 
parameters $\mathbf p(t)$ as
\begin{equation}
    \frac{\partial}{\partial p_i^*} \frac{\langle \bar \Psi(t)|\hat H(t) - i\frac{\partial}{\partial t} | \bar \Psi(t) \rangle}{\langle \bar \Psi(t) | \bar \Psi(t) \rangle}=0,
\label{eq:frenkel}
\end{equation}
where the parameters can now in general be complex. To apply this principle to linear order 
in $\gamma$, the wave function is expanded to second order in $\delta \mathbf p(t)$ around $\mathbf p_0$:
\begin{equation}
    |\bar \Psi(t) \rangle = |\bar \Psi_0 \rangle + \sum_i \delta p_i(t) |\bar \Psi_i \rangle + \frac{1}{2} \sum_{ij} \delta p_i(t) \delta p_j(t) |\bar \Psi_{ij} \rangle\,,
\end{equation}
with $| \bar \Psi_{i} \rangle =| \partial_i \bar \Psi \rangle$ and $| \bar \Psi_{ij} \rangle =| \partial_i \partial_j \bar \Psi \rangle$
computed at the parameters $\mathbf p_0$. These can be explicitly written as
\begin{eqnarray}
    |\bar \Psi_{i} \rangle\ &=& | \Psi_{i} \rangle - \langle \Psi_0 | \Psi_i \rangle |\Psi_0 \rangle \nonumber\\
    |\bar \Psi_{ij} \rangle &=& | \Psi_{ij} \rangle - \langle \Psi_0 | \Psi_i \rangle |\Psi_j\rangle - \langle \Psi_0 | \Psi_j \rangle |\Psi_i\rangle \nonumber\\
    &+&\left( 2\langle \Psi_0 | \Psi_i \rangle \langle \Psi_0 | \Psi_j \rangle - \langle \Psi_0 | \Psi_{ij} \rangle \right) | \Psi_0 \rangle\,,
\end{eqnarray}
where we use the same notation as above for the derivatives of $\Psi$, namely, $| \Psi_{i} \rangle =| \partial_i \Psi \rangle$ and $| \Psi_{ij} \rangle =| \partial_i \partial_j \Psi \rangle$.
Inserting this wave function in Eq.~\ref{eq:frenkel} and keeping only the first-order terms in $\delta \mathbf p(t)$, in the limit of $\gamma \rightarrow 0$, one obtains
\begin{equation}
    \mathbf A\, \delta \mathbf p(t) + \mathbf B\, \delta \mathbf p(t)^* = i\bar{\mathbf S}\, \frac{\partial \delta \mathbf p(t)}{\partial t}\,,
\end{equation}
with the matrix elements $A_{ij} =  \langle \bar \Psi_i | \hat H -E_0 | \bar \Psi_j\rangle = \bar H_{ij} - E_0 \bar S_{ij}$ and $B_{ij} = \langle \bar \Psi_{ij} | \hat H | \Psi_0 \rangle$. If we search for an oscillatory solution,  
\begin{equation}
    \delta \mathbf p(t)=\mathbf X_n e^{-i\omega_n t} + \mathbf Y_n^* e^{i\omega_n t}\,,
\end{equation}
with $\omega_n$ an excitation energy and $\mathbf X_n$ and $\mathbf Y_n$ the response vectors, we obtain
the well-known linear-response equations, here formulated as a non-Hermitian generalized eigenvalue equation,
\begin{equation}
    \begin{pmatrix}
        \mathbf A & \mathbf B \\
        \ \mathbf B^* & \ \mathbf A^*
    \end{pmatrix}
    \begin{pmatrix}
        \mathbf X_n \\
        \mathbf Y_n
    \end{pmatrix}
    = \omega_n
    \begin{pmatrix}
        \bar{\mathbf S} & \mathbf 0 \\
        \mathbf 0 & -\bar{\mathbf S}^*
    \end{pmatrix}
    \begin{pmatrix}
        \mathbf X_n \\
        \mathbf Y_n
    \end{pmatrix}\,.
\end{equation}
Neglecting $\mathbf B$ leads to the Tamm-Dancoff approximation,
\begin{equation}
    \mathbf A \mathbf X_n = \omega_n \bar{\mathbf S} \mathbf X_n\,,
\end{equation}
which is equivalent to the generalized eigenvalue equations of the linear method (Eq.~\ref{eq:lm-eigenvalue}) 
for an optimized ground-state wave function (when the gradients of the energy are therefore zero):
\begin{equation}
    \bar{\mathbf H} \mathbf X_n = (\omega_n+E_0) \bar{\mathbf S} \mathbf X_n\,.
\end{equation}
The energy $E_0$ is the variational energy, $E_\mathrm v$, for the optimized ground state.
Therefore, upon optimization of the ground-state wave function in the linear method, we can simply use the higher
roots resulting from the diagonalization of the equations to estimate the excitation energies as $\omega_n = (E_\text{lin})_n - E_0$ 
together with the oscillator strengths~\cite{mussard_chapter_2018}.

The time-dependent linear-response VMC approach has so far only been applied to
the excitations of the beryllium atom within the Tamm-Dancoff approximation and
with a simple single-determinant Jastrow-Slater wave
function~\cite{mussard_chapter_2018}. These calculations represent an
interesting proof of principle that multiple excitations of different space and
spin symmetry can be readily obtained after optimizing the ground-state wave
function. A systematic investigation with multi-configurational wave functions
is needed to fully access the quality of the approach, also beyond the
Tamm-Dancoff approximation.

\subsection{Variance-based methods}
\label{sec:varmin}

Variance minimization is a different approach to optimize the wave function compared to the methods described 
so far, which allows in principle to optimize excited states in a state-specific fashion. The target quantity for the
minimization is the variance of the energy:
\begin{equation}\label{eq:variance-minimization}
	\sigma_\mathrm v^2 = \frac{\langle \Psi | (\hat H - E_\mathrm v)^2 | \Psi\rangle}{\langle \Psi | \Psi \rangle}
        = \frac{\int \Psi^2(\mathbf R)(E_\mathrm L(\mathbf R)-E_\mathrm v)^2\mathrm d \mathbf R}{\int \Psi^2(\mathbf R) \mathrm d \mathbf R}\,.
\end{equation}
While the (global) minimum of the variational energy is only obtained for the
ground state, the variance has a known minimum of zero for each eigenstate of
the Hamiltonian. The optimization of the variance can be performed using either
a Newton approach with an approximate expression of the Hessian of the
variance~\cite{Umrigar2005} or reformulated as a generalized eigenvalue problem,
namely, a linear method for the optimization of the variance~\cite{Toulouse2008}.  For excited
states, the initial guess of the trial wave function might however be very
important to select a specific state and ensure that the minimization of the variance
leads to the correct local minimum.

More robust state-specific variational principles for excited states can be formulated so that the optimization 
of the excited state yields a minimum close to an initial target energy.  A simple possibility is to substitute 
the wave-function-dependent average energy in $\sigma_\mathrm v$ with a guess value $\omega$ as in
\begin{equation}
    \sigma^2_\omega = \frac{\langle \Psi | (\hat H - \omega)^2 | \Psi \rangle}{\langle \Psi | \Psi \rangle} \equiv (E_\mathrm v -\omega)^2+\sigma_\mathrm v^2\,,
\end{equation}
as it was done in the early applications of variance minimization on a fixed Monte Carlo sample~\cite{Umrigar1988}, where $\omega$
was chosen equal to a target value at the beginning of the optimization and then adjusted to the current best energy. If one updates $\omega$ in this manner, minimizing $\sigma_\omega$ is equivalent to minimizing $\sigma_\mathrm v$.

Alternatively, minimization of the variance can also be achieved by optimizing the recently proposed functional $\Omega$ defined as
\begin{equation}
\Omega = \frac{\langle \Psi | \omega - \hat H | \Psi \rangle}{\langle \Psi | (\omega - \hat H)^2 | \Psi \rangle} = \frac{\omega - E_\mathrm v}{(\omega - E_\mathrm v)^2 + \sigma_\mathrm v}\,,
\end{equation}
where $\omega$ is adjusted during the optimization to be equal to the current value of $E_\mathrm v-\sigma_\mathrm v$~\cite{zhao_efficient_2016,shea_size_2017}.  While the functional has formally a minimum for a state with energy directly above $\omega+\sigma_\mathrm v$, 
keeping $\omega$ fixed would lead to lack of size consistency in the variational principle~\cite{shea_size_2017}. Therefore, after some initial iterations, $\omega$ is gradually varied to match the current value of $E_\mathrm v-\sigma_\mathrm v$ required to achieve variance minimization. 
As in the case of the energy and the variance, this functional
can be optimized in VMC through a generalization of the linear method~\cite{zhao_efficient_2016,flores_excited_2018}. 

\section{Applications to excited states of molecular systems}

The status of excited-state quantum Monte Carlo calculations closely parallels
the me\-tho\-dological developments that have characterized the last decade as
we have outlined above in the context of wave function optimization. Since the
early applications to excited states, QMC methods were mainly employed as a tool
to compute vertical excitation energies and validate results of more approximate
methods.  Input from other -- sometimes much less accurate -- quantum chemical
approaches was however then used for the construction of the wave function, whose
determinant component was generally not optimized in the presence of the Jastrow
factor. The success of the calculation was therefore often heavily relying on the
ability of DMC to overcome possible shortcomings of the chosen trial wave
function. This must be contrasted to the recent situation of VMC having matured
to a fully self-consistent method as regards the wave function \textit{and} the
geometry with a rich ecosystem of tools ranging from basis sets and
pseudopotentials to multi-scale formulations.

One of the first QMC computations of two states of the same symmetry was carried
out for the H$_2$ molecule~\cite{grimes_quantum_1986}: the wave function was
obtained from a multi-reference calculation and DMC was shown to be able in this 
case to correct for the wave function bias.  Over the subsequent years, a number of
studies of vertical excitation energies were carried out with this basic
recipe, namely, performing DMC calculations on a given simple wave
function obtained at a lower level of theory~\cite{williamson_quantum_2002,
hood_quantum_2003, puzder_computational_2003,el_akramine_quantum_2003,
aspuru-guzik_quantum_2004, drummond_electron_2005, bande_rydberg_2006,
vincent_quantum_2007,tiago_neutral_2008, marsusi_comparison_2011}.  
Some of these early excited-state calculations were in fact pioneering
as they were applied to remarkably large systems such as silicon and
carbon nanoclusters with more than hundred
atoms~\cite{williamson_quantum_2002,puzder_computational_2003,drummond_electron_2005}.
Given the size of the systems, the choice of excited-state wave function was then
very simple and consisted of a single determinant correlated with a Jastrow factor
and constructed with the HOMO and LUMO orbitals from a density functional theory
(DFT) calculation.  Nevertheless, the resulting DMC excitation energies clearly
represented an improvement on the time-dependent DFT values and captured much of
the qualitative physics of the problem.  Even though doubts on the validity of this simplistic
recipe~\cite{tiago_neutral_2008,marsusi_comparison_2011} led researchers to investigate the
use of orbitals and pseudopotentials obtained with different density
functionals~\cite{drummond_electron_2005} as well as a multi-determinant
description~\cite{marsusi_comparison_2011}, a rather heuristic approach erring
on the side of computational saving characterized excited-state calculations in
this earlier period.

\begin{figure}
    \includegraphics[width=\linewidth]{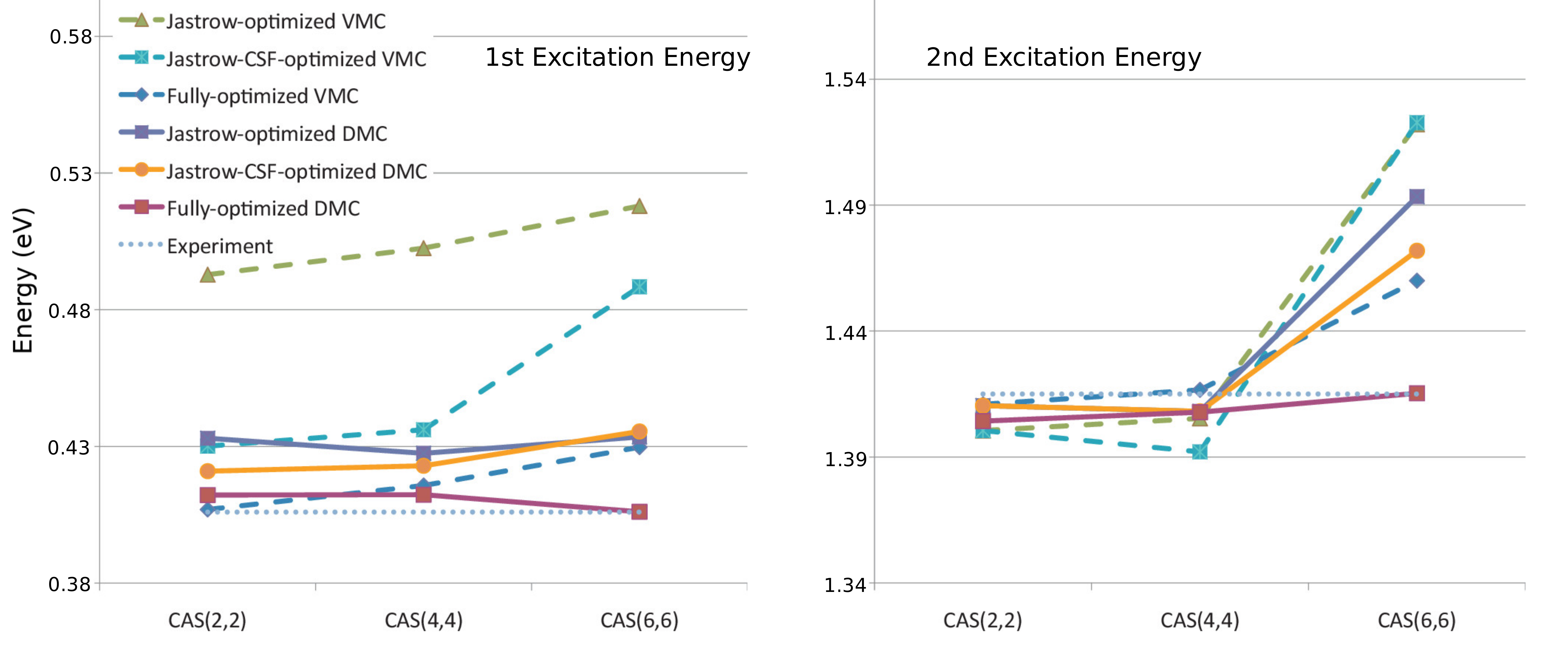}
    \caption{Convergence of VMC and DMC adiabatic excitation energies for the first and second excited states of methylene with increasing CAS size. Three levels of optimization have been used for the wave function: Jastrow, Jastrow and linear coefficients, and all parameters. Adapted from Ref.~\cite{zimmerman_excited_2009}.}
    \label{fig:es-cas}
\end{figure}

More recently, the development of algorithms for wave function optimizations in a 
state-specific or state-average fashion has
allowed us to better understand the proper ingredients in an excited-state QMC
calculation through the study of simple but challenging molecules. In
particular, it has become apparent that large improvements in the accuracy of
both VMC and DMC excited states can be achieved by optimizing the determinantal
component in the presence of the Jastrow factor at the VMC
level~\cite{schautz_optimized_2004,schautz_excitations_2004, filippi_absorption_2009, zimmerman_excited_2009, send_electronic_2011,blunt_excited-state_2019}. For instance, accurate excitations for
low-lying states of ethene can only be obtained if the orbitals
derived from a complete-active-space self-consistent-field (CASSCF) calculation
are reoptimized in the presence of the Jastrow factor to remove spurious
valence-Rydberg mixing in the final DMC energies~\cite{schautz_optimized_2004}. 
The analysis of various small organic molecules reveals that the use of
simplistic wave functions such as a HOMO-LUMO Hartree-Fock or a
configuration-interaction-singles ansatz may lead to significant errors also in
DMC~\cite{schautz_excitations_2004}.  An investigation of the ground and excited
states of methylene shows that the optimization of all variational parameters
reduces the dependency of both VMC and DMC on the size of the active space
employed for the trial wave function as shown in Fig.~\ref{fig:es-cas}~\cite{zimmerman_excited_2009}. In general,
while a minimal requirement is to optimize the linear coefficients
together with the Jastrow factor, the optimization of the orbitals is highly
recommended, especially if one employs a truncated expansion in computing the
excitation energies. Furthermore, evidence has been given that the optimization of
excited-state geometries requires the optimization of all wave
function parameters in order to obtain accurate gradients and, consequently,
geometries~\cite{valsson_photoisomerization_2010}. Finally, by construction,
linear-response VMC depends strongly on the quality of the ground-state wave
function for the description of excited states and, therefore benefits
considerably from orbital optimization in the ground
state~\cite{mussard_chapter_2018}.

While these and other examples of excited-state QMC calculations clearly
illustrate the importance of using wave functions with an adequate description of static
correlation and consistently optimized in VMC, they also demonstrate
the robustness of QMC approaches and some of their advantages with respect to
standard multi-configurational methods. In particular, the VMC and DMC
excitations are well converged already when very few determinants of a CAS
expansion are kept in the determinantal component of the wave
function~\cite{filippi_absorption_2009,send_electronic_2011}. Furthermore, the
demands on the size of the basis set are also less severe and one can obtain
converged excitation energies with rather small basis
sets~\cite{valsson_photoisomerization_2010, Zulfikri2016, blunt_excited-state_2019}. We note that most of the recent QMC calculations for
excited states have attempted to achieve a balanced static description of the
states of interest either by employing a CAS in the determinantal component or a
truncated multi-reference ansatz where one keeps the union of the configuration
state functions resulting from an appropriate truncation scheme (e.g.\ the sum of
the squared coefficients being similar for all states). Interestingly, matching
the variance of the states has recently been put forward as a more robust
approach to achieve a balanced treatment of the states in the computation of
excitation energies~\cite{Robinson2017,flores_excited_2018}.

\begin{figure}[t]
	\centering
	\includegraphics[width=0.8\linewidth]{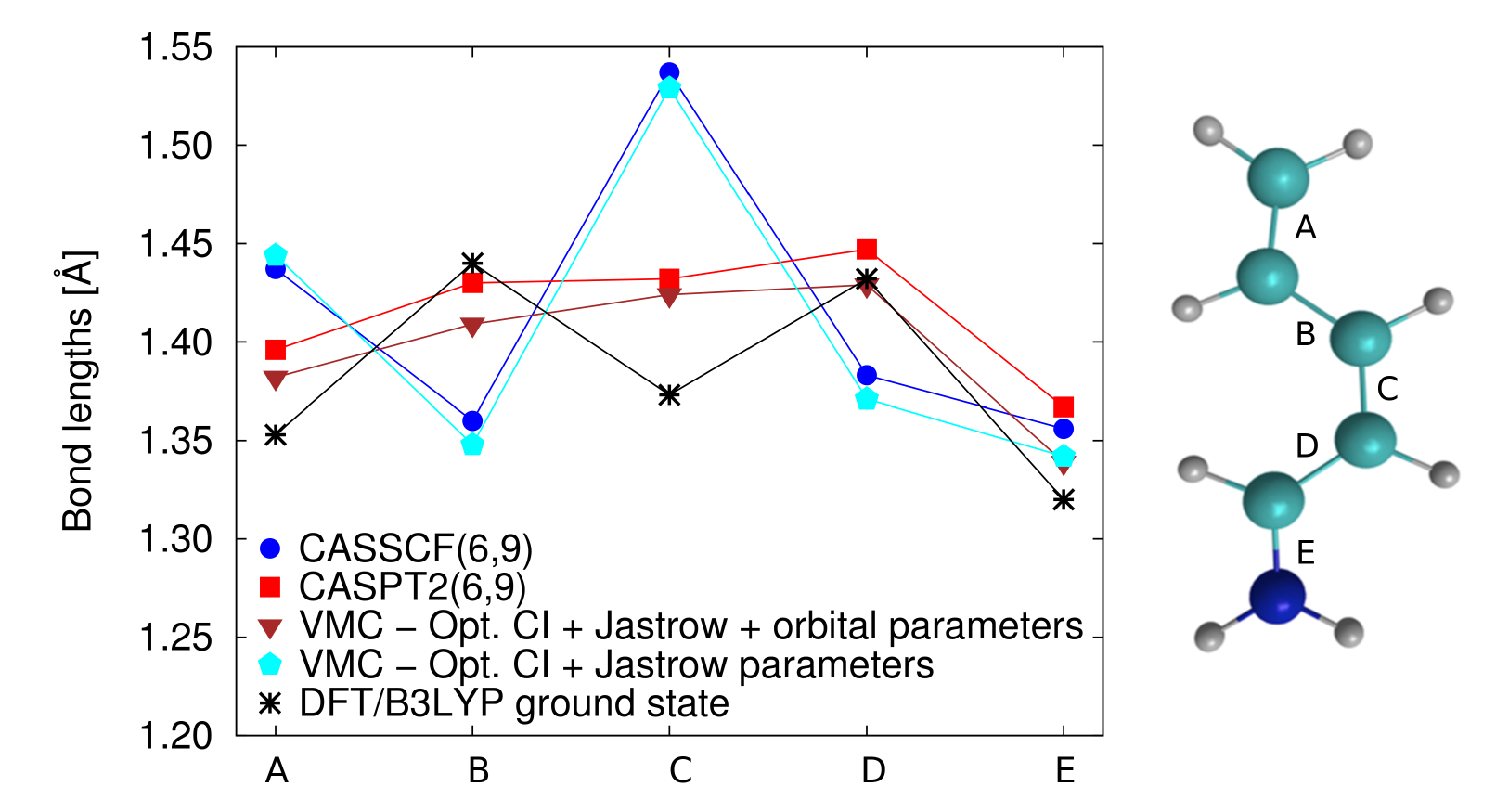}
	\caption{Planar excited-state geometry of a retinal protonated Schiff base model optimized with CASSCF, second-order perturbation theory (CASPT2), and VMC. Adapted from Ref.~\cite{valsson_photoisomerization_2010}.}
	\label{fig:geometries}
\end{figure}
The ability to optimize geometries even in the ground states has been a very
recent achievement for QMC methods, so most QMC calculations also outside the
Franck-Condon region have been performed on geometries obtained at a different
level of theory~\cite{schautz_excitations_2004,cordova_troubleshooting_2007, 
tapavicza_mixed_2008,zimmerman_excited_2009,dubecky_ground_2010,marsusi_comparison_2011,flores_excited_2018}. 
Nevertheless, these investigations have led to very promising
results, showing interesting prospects for the application of QMC to
geometry relaxations in the excited state, where most quantum chemical methods
either lack the required accuracy or are computationally prohibitive due to
their scaling with system size. For example, QMC was successfully employed to
assess the accuracy of various time-dependent DFT methods in describing the
photochemisty of oxirane through exploration of multiple excited-state potential
energy surfaces, also in proximity of conical intersection
regions~\cite{cordova_troubleshooting_2007, tapavicza_mixed_2008}. Another
application
demonstrating the very good performance of DMC was the study of different
conformers of azobenzene in the ground and excited
states~\cite{dubecky_ground_2010}. To the best of our knowledge, to date, the
only few attempts to optimize an excited-state
geometry via QMC gradients are our studies of the retinal protonated Schiff base
model~\cite{valsson_photoisomerization_2010,Zulfikri2016} and benchmark
calculations on small organic molecules in the gas phase~\cite{Guareschi2013}
and in a polarizable continuum model~\cite{guareschi_solvent_2014}. As shown for
the retinal minimal model in Fig.~\ref{fig:geometries}, the results are very
encouraging as they demonstrate that the QMC structures relaxed in the excited
state are in very good agreement with other highly-correlated approaches.
As already mentioned, the VMC gradients are
sensitive to the quality of the wave function and the orbitals must be
reoptimized in VMC to obtain accurate results. Tests also indicate
that the use of DMC gradients is not necessary as DMC cannot compensate for the
use of an inaccurate wave function while it yields comparable results to VMC
when the fully optimized wave function is employed.

Finally, we mention the recent developments of multi-scale methods in
combination with QMC calculations for excited states. Multi-scale approaches are
particularly relevant for the description of photoactive processes that can be
traced back to a region with a limited number of atoms, examples being a
chromophore in a protein or a solute in a solvent.  While this locality enables
us to treat the photoactive region quantum mechanically, excited-state
properties can be especially sensitive to the environment (e.g.\ the polarity of
a solvent or nearby residues of a protein), which cannot therefore be neglected but
are often treated at lower level of theory.
Multi-scale approaches are well-established in traditional quantum chemistry but
represent a relatively new area of research in the context of QMC. First steps
in this direction were made by combining VMC with a continuum solvent model,
namely, the polarizable continuum model (PCM)~\cite{Amovilli2008}. The approach
was used to investigate solvent effects on the vertical excitation energies of
acrolein~\cite{floris_electronic_2014} and on the optimal excited-state geometries of a number of small
organic molecules~\cite{guareschi_solvent_2014}. A
notable advantage of VMC/PCM is that the interaction between the polarizable
embedding and the solute is described self-consistently at the same level of
theory. This stands for instance in contrast to perturbation approaches which
include the interaction with the environment obtained self-consistently only at
the zero-order level (e.g.\ CASSCF).

\begin{figure}[htb]
    \includegraphics[width=\linewidth]{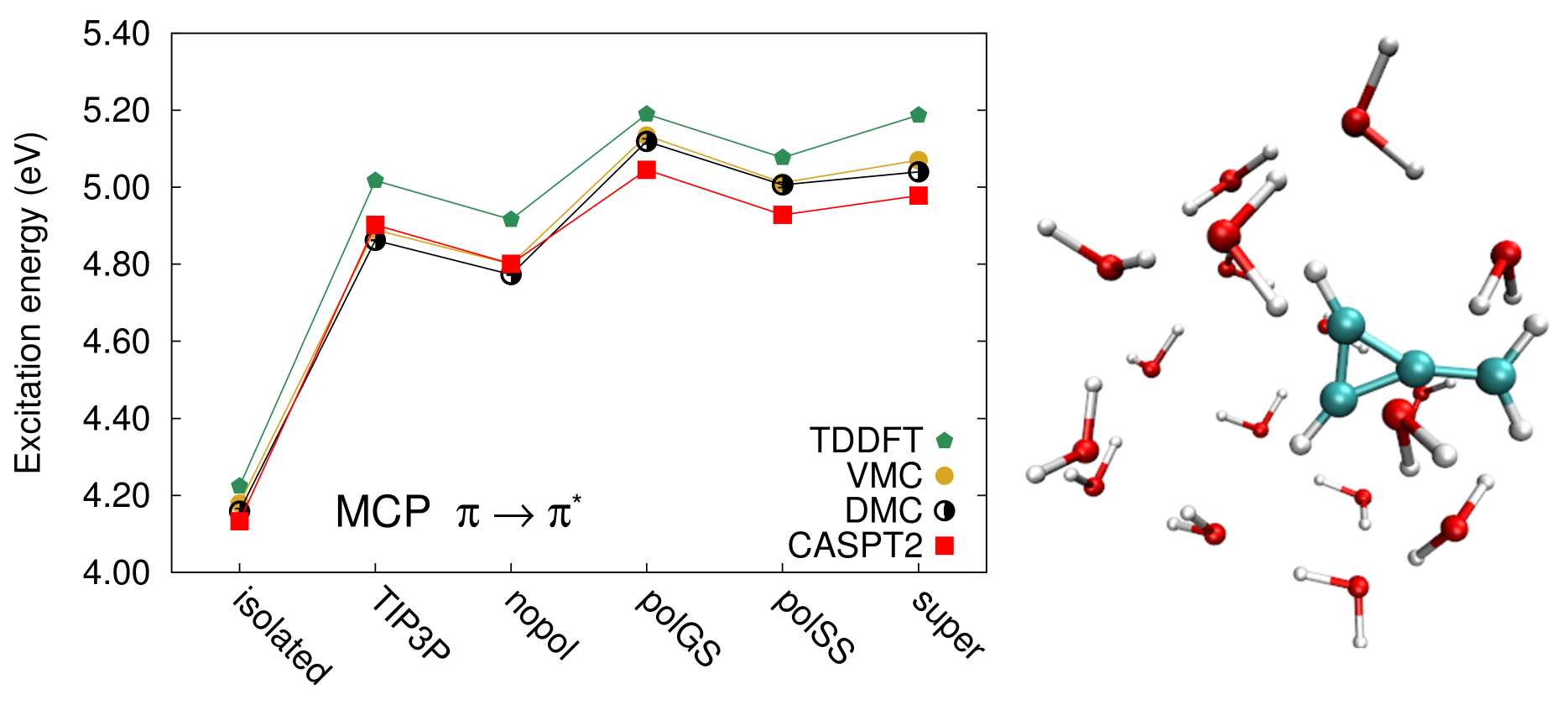}
    \caption{Energies of the $\pi\rightarrow \pi^*$ excitation of methylenecyclopropene in vacuum and embedded in a cluster of water molecules, computed with time-dependent DFT, CASPT2, VMC, and DMC. The water molecules are described with a static TIP3P force field and a MMpol approach with no polarization (nopol), with induced ground-state (polGS) and state-specific (polSS) dipoles, and in a supermolecular calculation (super). Adapted from Ref.~\cite{guareschi_introducing_2016}.}
    \label{fig:qmmmpol}
\end{figure}

To achieve a more realistic description of the environment, a static molecular
mechanics environment coupled via electrostatic interactions with the VMC or DMC
chromophore was used to describe the absorption properties of the green
fluorescent protein and rhodopsin~\cite{filippi_bathochromic_2012, valsson_rhodopsin_2013}. The limitations of such a non-polarizable embedding
scheme led to further developments, replacing the static description with a
polarizable one in a so-called VMC/MMpol approach~\cite{guareschi_introducing_2016}. The force field consists of static
partial charges located at the positions of the atoms as well as atomic
polarizibilities. These are used to compute induced dipoles in equilibrium with
the embedded system at the level of VMC, which are then kept fixed in subsequent
DMC calculations. The computation of the QMC excitation energies can be carried
out for two polarization schemes: either the induced dipoles are determined for
the ground state and used also for the excited state (polGS), or the excitation
energy is computed in a state-specific manner as the difference between the
ground- and excited-state energies both obtained self-consistently in
equilibrium with the respective induced dipoles (polSS).  As illustrated in
Fig.~\ref{fig:qmmmpol}, the vertical excitation energies of small molecules in
water clusters depend strongly on the sophistication of the embedding scheme.
Only the polarizable force field with two separate sets of induced dipoles for
the ground and excited states leads to a very good agreement with the
supermolecular excitation. Again, the QMC results agree with complete-active-space 
second-order perturbation theory (CASPT2) which
however is found to be rather sensitive to the choice of the active space. The
most sophisticated QMC embedding scheme has so far been realized using a
wave-function-in-DFT method and including differential polarization effects
through state-specific embedding potentials~\cite{daday_wavefunction_2014}. As
in the case of the polSS approach, the use of different potentials in the ground
and excited state is particularly important for excitations which involve large
polarization effects due to a considerable rearrangement of the electron density
upon excitation.

\section{Alternatives to diffusion Monte Carlo}

In some of the applications we presented, VMC has been shown sufficient to provide accurate 
excited-state properties without the need to perform a DMC calculation. The reason is that 
the burden and complexity of the problem have now been moved from the DMC projection to the
construction and optimization at the VMC level of sophisticated wave functions with many parameters. It is therefore 
natural to ask if there are alternatives to DMC, which do not require us
to build and optimize complicated many-body wave functions.

The Cerperley-Bernu method~\cite{Ceperley1988} can in principle be used to compute the lowest-energy
eigenstates and the corresponding relevant matrix elements by constructing a set of many-body basis
states and improving upon them through the application of the imaginary-time projection operator also
used in DMC. The Hamiltonian and overlap matrices are computed on these improved basis states during
the projection and the eigenvalues and eigenstates are then obtained by solving this
generalized eigenvalue problem. The method requires however that the fixed-node constraint is
relaxed during the projection, and therefore amounts to an expensive ``nodal-release'' approach.  The
approach has been successfully applied to the computation of low-lying excitations of bosonic 
systems~\cite{Nightingale2001,Ceperley1988}
and has also been used to compute tens of excited states of the fermionic, high-pressure
liquid hydrogen in order to estimate its electrical conductivity~\cite{Lin2009}.

If we move beyond a continuum formulation of QMC, the auxiliary-field quantum Monte Carlo (AFQMC) 
method by Zhang and coworkers~\cite{Zhang2003} represents a very distinct, feasible alternative 
to DMC. In this approach, the random walk is in a
space of single-particle Slater determinants, which are subject to a fluctuating external potential.
The fermion-sign problem appears here in the form of a phase problem and is approximately eliminated
by requiring that the phase of the determinant remains close to the phase of a trial wave
function.  The method is more expensive than DMC but has been applied to a variety of
molecular and extended systems (mainly in the ground state) and appears to be less plagued by the phase constraint
as compared to the effect of the fixed-node approximation in DMC. 
A very recent review of AFQMC and
its applications also to excited states can be found in Ref.~\cite{Zhang2018}. 

Finally, we should mention another QMC approach in determinantal space, namely, the full configuration interaction 
quantum Monte Carlo method~\cite{Booth2009,Cleland2010} where a stochastic approach is used to select the important 
determinants in a full configuration interaction expansion.  The method has been described in Chapter X together with
its extension to excited states.

\bibliographystyle{ieeetr}
\bibliography{qmc}
\end{document}